\documentclass[sigconf]{acmart}

\AtBeginDocument{}

\copyrightyear{2024}
\acmYear{2024}
\setcopyright{rightsretained}
\acmConference[CHI '24]{Proceedings of the CHI Conference on Human Factors in Computing Systems}{May 11--16, 2024}{Honolulu, HI, USA}
\acmBooktitle{Proceedings of the CHI Conference on Human Factors in Computing Systems (CHI '24), May 11--16, 2024, Honolulu, HI, USA}
\acmDOI{10.1145/3613904.3642083}
\acmISBN{979-8-4007-0330-0/24/05}

\usepackage{xspace}
\usepackage{calc}
\usepackage{stfloats}
\begin{document}
\newcommand{\todo}[1]{}
\renewcommand{\todo}[1]{{\color{red} \textbf{TODO: {#1}}}}

\newcommand{\fix}[1]{}
\renewcommand{\fix}[1]{{\color{orange} \textbf{FIXME: {#1}}}}

\newcommand{\cit}[1]{}
\renewcommand{\cit}[1]{{\color{green} \textbf{CITEME: {#1}}}}

\newcommand{\system}{pARam\xspace}
\newcommand{\concept}{in-situ interaction with parametric designs\xspace}
\newcommand{\param}{\system}

\newcommand{\systemEmph}{\emph{\system}\xspace}
\newcommand{\systemEmphSp}{\systemEmph\xspace}

\newcommand{\tv}{Thingiverse\xspace}
\newcommand{\mmf}{MyMiniFactory\xspace}
\newcommand{\pa}{Printables\xspace}

\newcommand{\fn}[2]{\footnote{\url{#1}, \mbox{Accessed:~{#2}}}}

\newcommand{\m}{\textit{M=}}
\newcommand{\sd}{\textit{SD=}}
\newcommand{\F}[3]{$F({#1},{#2})={#3}$}
\newcommand{\p}{\textit{p=}}
\newcommand{\pminor}{\textit{p$<$}}
\newcommand{\chisq}{$\chi^2$}
\newcommand{\avg}[1]{$avg.={#1}$}

\newcommand{\mrank}{\textit{$M_{rank}$=}}
\newcommand{\pref}[2]{($desktop: {#1}\%, \system: {#2}\%$)}
\newcommand{\preferr}[2]{(desktop: {#1}, \system: {#2})}
\newcommand{\iquote}[2]{\textit{``{#1}''}~({#2})}
\newcommand{\dirquote}[1]{\textit{``{#1}''}}

\definecolor{expert}{RGB}{129, 122, 185}
\definecolor{nonexpert}{RGB}{229, 174, 163}

\renewcommand*{\figureautorefname}{Fig.}
\renewcommand*{\subsectionautorefname}{section}

 \title[pARam]{pARam: Leveraging Parametric Design in Extended Reality to Support the Personalization of Artifacts for Personal Fabrication}

\author{Evgeny Stemasov}
\orcid{0000-0002-3748-6441}
\email{evgeny.stemasov@uni-ulm.de}
\affiliation{\institution{Ulm University}
  \city{Ulm}
  \country{Germany}
}

\author{Simon Demharter}
\orcid{0000-0002-8768-2942}
\email{simon.demharter@uni-ulm.de}
\authornote{Both authors contributed equally to this research.}
\affiliation{\institution{Ulm University}
  \city{Ulm}
  \country{Germany}
}

\author{Max Rädler}
\orcid{0000-0002-5413-2637}
\email{max.raedler@uni-ulm.de}
\authornotemark[1]
\affiliation{\institution{Ulm University}
  \city{Ulm}
  \country{Germany}
}

\author{Jan Gugenheimer}
\orcid{0000-0002-6466-3845}
\email{jan.gugenheimer@tu-darmstadt.de}
\affiliation{\institution{TU-Darmstadt}
  \city{Darmstadt}
  \country{Germany}
}
\affiliation{\institution{Institut Polytechnique de Paris}
  \city{Paris}
  \country{France}
}

\author{Enrico Rukzio}
\orcid{0000-0002-4213-2226}
\email{enrico.rukzio@uni-ulm.de}
\affiliation{\institution{Ulm University}
  \city{Ulm}
  \country{Germany}
}

\renewcommand{\shortauthors}{Stemasov, Demharter, Rädler, Gugenheimer, Rukzio}

\begin{abstract}
    Extended Reality (XR) allows in-situ previewing of designs to be manufactured through Personal Fabrication (PF). These in-situ interactions exhibit advantages for PF, like incorporating the environment into the design process. 
However, design-for-fabrication in XR often happens through either highly complex 3D-modeling or is reduced to rudimentary adaptations of crowd-sourced models. We present pARam, a tool combining parametric designs (PDs) and XR, enabling in-situ configuration of artifacts for PF.
In contrast to modeling- or search-focused approaches, pARam supports customization through embodied and practical inputs (e.g., gestures, recommendations) and evaluation (e.g., lighting estimation) without demanding complex 3D-modeling skills.
We implemented pARam for HoloLens 2 and evaluated it ($n=20$), comparing XR and desktop conditions. Users succeeded in choosing context-related parameters and took their environment into account for their configuration using pARam. We reflect on the prospects and challenges of PDs in XR to streamline complex design methods for PF while retaining suitable expressivity.

 \end{abstract}



\begin{teaserfigure}
\centering
\includegraphics[width=1\linewidth]{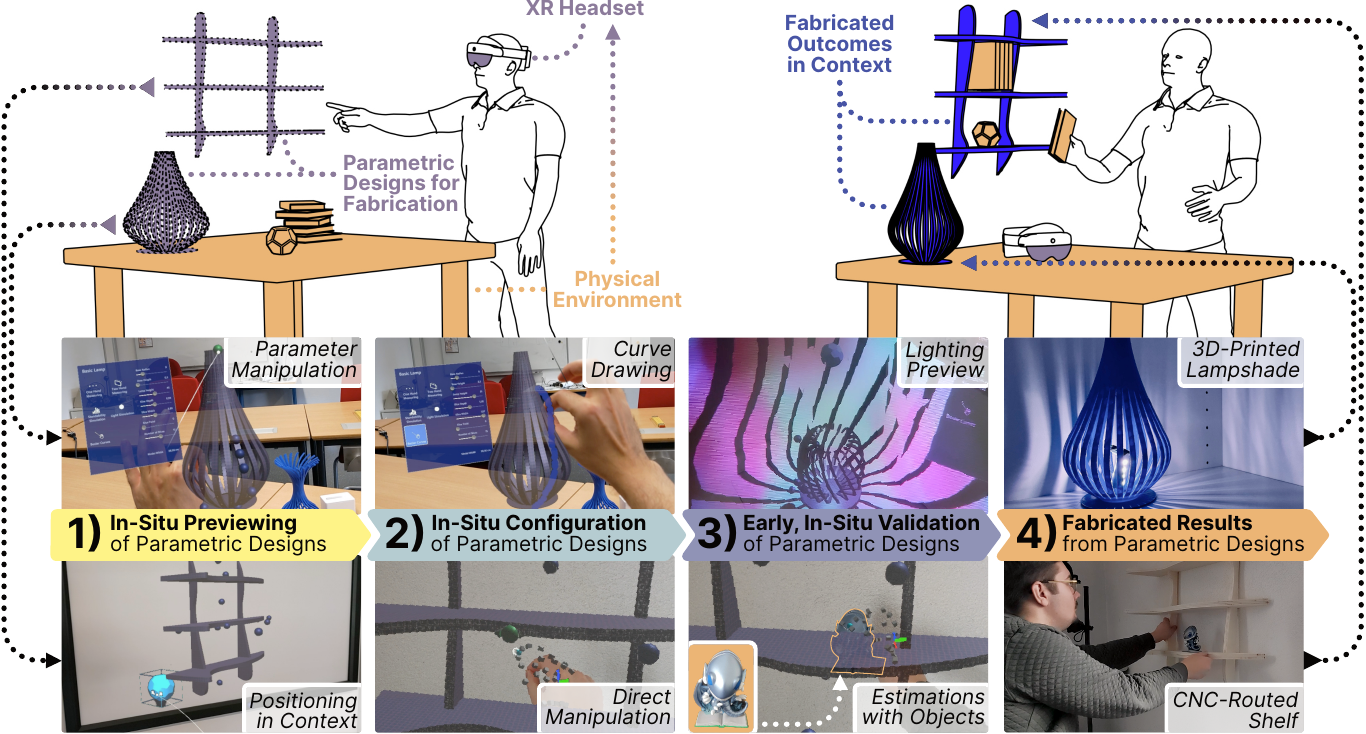}
  \caption{\system enables \concept through extended reality. It allows users to position and preview fabricatable parametric designs in the context they will be used in after fabrication~(1). Users then configure the designs to their liking and to their contextual requirements, leveraging support mechanisms like mid-air sketching or recommendations~(2), and validate them with respect to existing objects or their environment (e.g., through lighting estimation) in-situ~(3). This allows users to achieve fitting results that are fabricated based on their in-situ customization~(4) without engaging in (3D-)modeling ``from scratch'' or being restricted to finished, immutable designs. }
  \Description{}
  \label{fig:teaser}
\end{teaserfigure}

\maketitle

Personal fabrication (PF) is within reach of a broad range of users: increasing numbers of people have access to the means of industry-grade manufacturing devices, such as 3D-printers or laser cutters~\cite{gershenfeldFabComingRevolution2005,motaRisePersonalFabrication2011}.
This access is facilitated through device ownership or infrastructure like makerspaces~\cite{soomroMakerspacesFosteringCreativity2023}, print centers~\cite{hudsonUnderstandingNewcomers3D2016}, or fablabs~\cite{garnierMakingHackingCoding2021,gershenfeldDesigningRealityHow2017}.
The means for personal fabrication empower end-users to create unique artifacts following their requirements and express themselves in a physical, tangible medium.
Two contrasting approaches to designing for fabrication exist: a) \textbf{modeling}, where users define most, if not all, details of a design~\cite{stemasovRoadUbiquitousPersonal2021}; and b) \textbf{retrieval}, where users search for finished models~\cite{alcockBarriersUsingCustomizing2016,stemasovRoadUbiquitousPersonal2021} in repositories like Printables or Thingiverse, which provide access to crowd-sourced designs~\cite{liangCustomizARFacilitatingInteractive2022}.
However, both approaches have their shortcomings.
3D-modeling for personal fabrication fundamentally requires users to invest time in learning and applying \emph{domain knowledge} from various disciplines, like design, engineering, or material science~\cite{stemasovRoadUbiquitousPersonal2021,hudsonUnderstandingNewcomers3D2016}.
Retrieval, on the other hand, is constrained by the available databases and suffers from hard-to-alter designs~\cite{alcockBarriersUsingCustomizing2016}.
A middle ground is found in the notion of \textbf{remixing}.
In the context of digital media~\cite{navasRemixTheoryAesthetics2014}, creative programming~\cite{hillCostCollaborationCode2013} but also in personal fabrication~\cite{oehlbergPatternsPhysicalDesign2015}, remixing allows creative practitioners to benefit from collaborative effort and avoid ``starting from scratch''~\cite{stemasovRoadUbiquitousPersonal2021}.
For personal fabrication, remixing is facilitated through interfaces to parametric designs: ``customizers''~\cite{liangCustomizARFacilitatingInteractive2022}.
This is particularly visible across model repositories like \tv\footnote{\url{https://www.thingiverse.com/app:22}, Accessed: 02.02.2024}~\cite{oehlbergPatternsPhysicalDesign2015}, where the presence of these easy-to-use interfaces to create adapted designs (remixes) sparked an increase of contributions~\cite{flathCopyTransformCombine2017}.
Parametric designs constrain the virtually infinite output space of a ``traditional'' design tool (e.g., Blender), to a limited set of valid designs for a specific object defined through combinations of parameters chosen by a ``primary author''~\cite{carpoSecondDigitalTurn2017}.
They, therefore, implicitly embed domain knowledge provided by the creator of the parametric design and may prevent errors~\cite{shugrinaFabFormsCustomizable2015,stemasovRoadUbiquitousPersonal2021}.
Comparable interactions--from a user perspective--are already present and are becoming increasingly relevant in consumer-facing systems like product configurators for furniture\footnote{e.g., \url{https://tylko.com/furniture-c/} or \url{https://www.pickawood.com/en/configurator/tables}, Accessed: 01.02.2024}, footwear\footnote{e.g., \url{https://www.bullfeet.com/en/custom-sport-shoes/personalize}, Accessed: 02.02.2024}, or other trinkets\footnote{e.g., \url{hhttps://www.shapeways.com/creator}, Accessed: 29.01.2024}.\\

Customizers constrain available input and output spaces. They balance complexity and expressivity by allowing users to start with a finished object definition, and tailor specific, relevant aspects of it to their liking.
However, they remain complex to successfully interact with, due to a set of issues: 
1) they \emph{operate ex-situ}, requiring transfers between physical context and design environment~\cite{ashbrookAugmentedFabricationCombining2016,mahapatraBarriersEndUserDesigners2019};
2) they often provide \emph{little to no support} in choosing specific values for parameters beyond constraints~\cite{shugrinaFabFormsCustomizable2015};
and 3) in the context of personal fabrication, the \emph{evaluation} of consequences is left to users and is often done through iteration, prototyping, and patching~\cite{teibrichPatchingPhysicalObjects2015}.
Ex-situ design means that novice users have to engage in (error-prone) measurement and transfer procedures, diminishing their chances of success~\cite{mahapatraBarriersEndUserDesigners2019}.
The absence of woven-in domain knowledge means that users have to explore and understand the effects of parameters on geometry, but also on subsequent process steps, like manufacturing (e.g., fabricability~\cite{shugrinaFabFormsCustomizable2015}) or usage (e.g., ergonomics~\cite{leeInteractiveSituatedGuidelines2018}). 
This is linked to the issue of evaluation, where a chosen parameter set may have non-trivial effects that users may only see after fabrication (e.g., stylistic consistency, lighting patterns), requiring them to iterate should they fail.\\

Based on formative explorations of product customizers, parameter types, and literature, we developed \system, a design tool to interact with, evaluate, and preview parametric designs for personal fabrication in-situ.
\system leverages the capabilities of Extended Reality (XR) devices to facilitate these interactions, which were previously bound to desktop devices (\emph{ex-situ}).
It was implemented as an application for the Microsoft HoloLens 2 and supports a total of 15 parametric designs.
\system allows users to preview~(\autoref{fig:teaser}.1) customize a parametric design in-situ through a 2D interface, direct manipulation, or gestures~(\autoref{fig:teaser}.2).
Lastly, before fabrication, users may assess their design \emph{in-situ}: using the environment mesh acquired by the HoloLens, they may evaluate the stability of objects, or changes to lighting and shadowing, for instance, for designing lampshades~(\autoref{fig:teaser}.3) through estimations provided by \system. 
The fabricated artifact~(\autoref{fig:teaser}.4) can then be placed in the context in which it was configured digitally. 
All aforementioned aspects bridge the disconnect between the space where parameters are defined (i.e., a design workstation), and the space where they become relevant and influential (i.e., the space they are fabricated for and used in).
To further support users in choosing fitting parameters, \system embeds domain knowledge beyond constraints (e.g, ergonomics-related suggestions).
All aforementioned aspects are meant to increase the expressivity of in-situ design tools that operate on existing designs, without demanding the complexity of highly complex, industry-grade 3D-modeling tools. \\

Through a user study ($n = 20$) that compared \system with a desktop-bound alternative, we found that users take their physical context into account while designing, and engage in more spatial interactions revolving around previewing and manipulating the model.
Users appreciated alternative means of defining parameter values, like using gestures to measure or draw curves. 
The participants preferred the desktop modality for functional objects, and preferred the AR variant for more aesthetic designs.
However, the in-situ and spatial nature of \system led users to engage more in ``eyeballing''~\cite{mahapatraBarriersEndUserDesigners2019} and approximations, which is appropriate for some, but not all, design tasks in the context of personal fabrication.\\

Our work contributes the following:
\begin{enumerate}
    \item \textbf{Presentation and discussion of the concept of \concept}, an emerging design approach for personal fabrication that leverages extended reality and parametric designs to enhance the expressivity of retrieval-oriented design tools.
    \item \textbf{Design, development, and evaluation ($n=20$) of \system}, an in-situ design tool to interact with parametric designs, enriched with input and validation steps enabled by extended reality, embodying the aforementioned concept.
\end{enumerate}  \section{Related Work}\label{sec:rw}
    \system is inspired by a set of overlapping research domains surrounding design and fabrication: parametric design tools, design tools for personal fabrication, and XR design tools. 
    \system is inspired by these works in several ways: It embraces the notions of simpler, easier-to-use design tools, situates relevant design effort in the usage context, and aims to embed domain knowledge in the system, instead of expecting users to acquire it over time.

    \paragraph{Definitions}\label{par:definitions}  
        The interaction with a parametric design can be viewed from two perspectives: the \textbf{creation} (i.e., authoring) of parametric objects and their \textbf{customization}.
        Mario Carpo describes this dichotomy as one \dirquote{[...] where the primary author designs a generic (parametric) object, and one or more secondary authors, or \textbf{interactors}, adjust and adapt some variable aspects of the original notation at will}~\cite{carpoSecondDigitalTurn2017}.
        This yields distinct roles of ``modeling'' and ``navigating''~\cite{costaEnablingParametricDesign2020}.
        Notably, design through a parametric model was described as a practice \dirquote{where design variations are effortless}~\cite{barrioshernandezThinkingParametricDesign2006}.
        In the scope of this work, we treat the ``interaction with a parametric design'' \emph{not} to be the act of authoring one (e.g., a designer creating customizable furniture in a tool like Grasshopper\fn{https://www.rhino3d.com/features/\#grasshopper}{30.01.2024}), but rather the \emph{act of customizing such a design to one's requirements and liking} (e.g., an end-user customizing one for their own home).
        The lens of end-users is particularly relevant for the space of personal fabrication, where users are handed the means to design and fabricate unique artifacts, but may lack the necessary domain knowledge~\cite{stemasovRoadUbiquitousPersonal2021,baudischPersonalFabrication2017} to do so ``from scratch''.
        
        Design can further be implemented as either an \emph{ex-situ} or an \emph{in-situ} process.
        In this context of personal fabrication, Ashbrook et al. defined \textbf{in-situ design} as a procedure where users design new artifacts \dirquote{with in its existing context}~\cite{ashbrookAugmentedFabricationCombining2016}.
        It is contrasted with \textbf{in-situ fabrication}, where the manufacturing of an object happens in conjunction with pre-existing ones~\cite{ashbrookAugmentedFabricationCombining2016,camposzamoraSPrintrInSituPersonal2022, stemasovEphemeralFabricationExploring2022}.
        \system focuses on \emph{in-situ design}, given that industrial fabrication capabilities may be easier to scale to unique, one-off designs in the context of mass-personalization~\cite{ozdemirDesignMethodologyMass2022}. 

    \subsection{Parametric Design Tools}
        \paragraph{Interacting with Parametric Designs}
            Various approaches to parametric design are present in domains of architecture or graphics, predating their use in personal fabrication~\cite{anderlParametricDesignIts1995}.
            Michel and Boubekeur presented a technical approach to enable direct, brush-like interaction with parametric shapes, which is also an outstanding example of an interactive technique for design~\cite{michelDAGAmendmentInverse2021}.
            Buyruk and ??a??da?? presented a mixed reality approach for architectural design that relies on robotic fabrication, introducing a framework unifying physical and digital components using a digital twin approach~\cite{buyrukInteractiveParametricDesign2022}.
Design space exploration was probed for generative design systems~\cite{matejkaDreamLensExploration2018}, artificial intelligence systems~\cite{urbandavisDesigningCoCreativeAI2021}, and situated in immersive environments~\cite{jenningsGeneratiVRSpatialInteractions2022,laoAttributeSpacesSupporting2021}.

            The aforementioned works have informed our design choices on fundamental interactions (e.g., sliders) but also considerations on validity (e.g., geometrically sound models).
            However, most approaches still separate design in disjoint design and usage environments--an aspect we seek to bridge with \system.

        \paragraph{Customizers in Personal Fabrication}
            Personal fabrication~\cite{baudischPersonalFabrication2017,motaRisePersonalFabrication2011}, in contrast to industrial manufacturing processes or graphics contexts, follows a different set of goals and constraints: not only are the outcomes physical, not digital, but, most importantly, the tools are also routinely used by non-experts~\cite{leeConsumerCreatorHow2017}. 
            This has manifested in various prototype systems that provide parametric design tools for the context of personal fabrication. 
            To design parametric designs themselves, tool support is similarly needed, for instance, in defining constraints in a structure to achieve valid and adaptive designs.
            This was addressed by Shugrina et al. with FabForms, an approach to ensuring that the outcomes from a customizer are not only geometrically valid but also manufacturable~\cite{shugrinaFabFormsCustomizable2015}.
            This weaves domain knowledge into a design or customization tool instead of expecting end-users to acquire and apply it.
            The creation of constraint-based parametric designs was explored through CODA by Veuskens et al., which provides tool support for constraint specification~\cite{veuskensCODADesignAssistant2021}.
The ``customizer'' on \tv\fn{https://www.thingiverse.com/app:22}{29.01.2024} is a popular example of an accessible parametric design configurator~\cite{flathCopyTransformCombine2017}.
            Its use has been studied extensively~\cite{oehlbergPatternsPhysicalDesign2015,voigtNotEveryRemix2018,acherCustomization3DPrinting2014}--while not impeccable, it, along with product customizers in commerce, is a viable entry point for novices who may not want to engage in prolonged learning processes (e.g., to learn 3D-modeling).
    
            \system is, fundamentally, a tool to \emph{interact} with a parametric design.
            The approach of parametric designs is well established in engineering and architecture~\cite{caetanoComputationalDesignArchitecture2020, stalsParametricModelingTools2021,carpoSecondDigitalTurn2017}. 
            Fundamentally, this approach constrains input and output spaces a user can leverage to design an object, turning this design process into a configuration.
            Based on this notion of ``configuring'' instead of ``modeling''~\cite{stemasovRoadUbiquitousPersonal2021}, parametric design tools became represented in everyday life in the form of various product configurators for consumer goods.
            \system aims to situate (i.e., in the context of use) and support (i.e., though validations) the design process of non-experts while focusing on a \emph{physical} output domain, namely manufactured artifacts.

        \subsection{Design Tools for Personal Fabrication}
            \paragraph{Novice-Oriented Design Tools}
                \emph{Design} is often conflated with (3D)-modeling, especially in the context of personal fabrication.
                However, other ways to acquire fitting designs are present in research.
Search and retrieval of models for fabrication has also been discussed in previous works focusing on previewing and remixing~\cite{stemasovMixMatchOmitting2020} or query formulation~\cite{stemasovShapeFindARExploringInSitu2022} in-situ.
                CustomizAR by Liang et al. is a tool to explore customizable designs on \tv, which further supported users with a tablet-based measurement functionality~\cite{liangCustomizARFacilitatingInteractive2022}. 
                The work is closely related to our work and \system.
                Liang et al. support the measurement of relevant dimensions but also the exploration of designs and further evaluate their camera-based measurement approach, confirming that it may provide sufficient precision.
                The authors further outlined how using established measurement tools may not prevent errors~\cite{liangCustomizARFacilitatingInteractive2022}, as measuring is a skill in and of its own~\cite{ramakersMeasurementPatternsUserOriented2023}.
                We aim to further explore and augment in-situ interactions with parametric designs, for instance, by providing recommendations and early validations.
                A comparably influential work for \system is BodyMeter by Lee et al.~\cite{leePosingActingInput2016}.
                The authors explored how the customization of furniture may be facilitated through spatial, embodied, and gestural interaction instead of detailed definition through modeling software.
                This exploration inspired both the \system's underlying concept and specific features, like the gesture+voice-based measurement component.
                
                All aforementioned works are relevant to \system, as they aim to balance low entry barriers with high, or at least sufficient-to-the-task expressivity.
                They leverage other approaches than ``modeling''~\cite{stemasovRoadUbiquitousPersonal2021}, yet allow users to find the designs they want or customize existing ones to their liking. 
                
                The aforementioned considerations are also present in various tools that enable the design for fabrication, while focusing on novice users, like Kyub~\cite{baudischKyub3DEditor2019}, FlatFitFab~\cite{mccraeFlatFitFabInteractiveModeling2014}, or application bridges like Blocks To CAD~\cite{lafreniereBlockstoCADCrossApplicationBridge2018}.
                Works by Follmer et al. presented ways to alter one's physical environment~\cite{follmerCopyCADRemixingPhysical2010} or make 3D-modeling accessible to children~\cite{follmerKidCADDigitallyRemixing2012}, both of which leverage \emph{remixing} over \emph{modeling} -- an approach we embrace with \system.
                SketchChair by Saul et al. provides a design environment for chairs, which, similarly to \system, supports users through simulation and ergonomics components~\cite{saulSketchChairAllinoneChair2011}. \system builds upon this premise in several ways: by situating the design process in the context of future use (in-situ), which avoids the need for ``reference geometry''~\cite{saulSketchChairAllinoneChair2011} and providing input support for a larger class of result objects~\cite{stemasovRoadUbiquitousPersonal2021}, beyond chairs.
                Ergonomics are a consideration present in various design tools, such as Body2Desk~\cite{leeInteractiveSituatedGuidelines2018}, which encapsulate domain knowledge~\cite{baudischPersonalFabrication2017} and support users in benefitting from it, without having to acquire it. 
                In contrast to \system, the work of Lee et al. focuses on desks as the core object~\cite{leeInteractiveSituatedGuidelines2018}, while \system aims to be an interface to parametric designs in general, along with a focus on interactions between artifact and context (in-situ), in addition to interactions between artifact and user~\cite{leeInteractiveSituatedGuidelines2018}.

            \paragraph{In-Situ Design for Fabrication} 
                In-situ design has emerged as a potential component of lower-effort design tools for personal fabrication.
                RefAR by Wu and Cheng~\cite{wuRefAR3DSketchBased2022} leverages depth sensing of a HoloLens to support a user's in-situ sketching process with features of the environment. 
                This allows replication of context features for a new design~\cite{wuRefAR3DSketchBased2022}.
                MixFab by Weichel et al.~\cite{weichelMixFabMixedrealityEnvironment2014} is a design environment enabling design in mixed reality, where physical objects can be brought into the space to use as features and references~\cite{weichelMixFabMixedrealityEnvironment2014}.
                Printy3D is an in-situ design tool that combines interactive projection and tangible building blocks, focused on the design of electronics enclosures made accessible to novice end-users--specifically children \cite{yungPrinty3DInsituTangible2018}.
                Tangible elements and XR have also been combined for headset-based design tools~\cite{stemasovBrickStARtEnablingInsitu2023}.

                \system embraces the notion of in-situ design for (personal) fabrication by situating the customization of parametric designs towards a user's context-driven needs.

            \paragraph{In-Situ Fabrication}
                The notion of ``in-situ'' actions can likewise be applied to the act of fabrication or manufacturing, where interacting objects found in a specific context, are included in the fabrication process.
                This includes approaches enabling mobile fabrication~\cite{roumenMobileFabrication2016,camposzamoraSPrintrInSituPersonal2022}, portable fabrication~\cite{peekPopfabCasePortable2017,quitmeyerWearableStudioPractice2015}, or wearable fabrication~\cite{stemasovEphemeralFabricationExploring2022}, each of them enabling in-situ fabrication activities. 
                RoMA by Peng et al.~\cite{pengRoMAInteractiveFabrication2018} is an example where design and fabrication happen in parallel, through a collaboration between a human user and a robotic arm, enabling a turn-taking process~\cite{pengRoMAInteractiveFabrication2018}.
                FusePrint is a smaller-scale setup, where physical objects are brought into a fabrication environment to replicate their features in complementary objects (e.g., pens brought in to support the design of a pen holder)~\cite{zhuFusePrintDIY5D2016}.
                In the context of architecture, IRoP by Mitterberger et al.~\cite{mitterbergerInteractiveRoboticPlastering2022} presents an approach to plastering that translates in-situ motion of designers to in-situ fabricated plastering~\cite{mitterbergerInteractiveRoboticPlastering2022}.
                The work of Jahn et al. is also situated in the architecture context and outlines a platform to move from 2D to 3D design processes and situate them in construction environments~\cite{jahnMakingMixedReality2018}.
                
                \system focuses on design in context but benefits from aspects like in-situ estimations to inform fabrication early on.
                This facilitates digital iteration, instead of physical prototyping.

        \subsection[Design in and for XR]{Design in and for Extended Reality}
            Extended Reality (XR) has also been both a tool and a target domain for design.
            Conceptually, there is a sizeable overlap between design tools for \emph{digital environments} (e.g., games) and design tools for \emph{physical artifacts}, both operating under the umbrella term of computer-aided design (CAD).

            \paragraph{Immersive Design}
                Remixed Reality by Lindlbauer and Wilson presented a set of interactions to alter a digitally acquired environment~\cite{lindlbauerRemixedRealityManipulating2018}. 
                This approach can be used to author altered versions of it--an outcome similarly targeted by \system, albeit through physical modifications.
                In an AR context, WindowShaping by Huo et al. presented interactions to use real-world elements for a design process~\cite{huoWindowShaping3DDesign2017} and Han et al. proposed methods to adapt AR interfaces to existing, real-world objects~\cite{hanBlendMRComputationalMethod2023}.
                This approach is manifested in the estimation functions of \system, where scanned aspects (e.g., the environment mesh) are used to support early evaluations and, by extension, design decisions.
                VRSketchIn provided users with a drawing tablet in VR, to author VR environments in-situ~\cite{dreyVRSketchInExploringDesign2020}, by providing \emph{support} to sketch one's designs.
                Concepts like ``Paramersive Design''~\cite{drogemullerEnvisioningParamersiveDesign2023}, similarly to \system, aim to leverage interaction paradigms known in architecture and enrich them with the means of XR~\cite{drogemullerEnvisioningParamersiveDesign2023}.

            \paragraph{Authoring XR Experiences}
                With immersive environments as the target, designers routinely engage in the creation of multimedial arrangements~\cite{stemasovImmersiveSamplingExploring2023}, which is a non-trivial task, especially for novices.
                Nebeling and Speicher provide a meta-perspective on design tools for extended reality, outlining several classes of tools, mapped by their fidelity and the required skill~\cite{nebelingTroubleAugmentedReality2018}. 
                Sketched Reality by Kaimoto et al. is a design approach that bridges digital sketches and physical output by coupling both dimensions, with each being able to affect the other~\cite{kaimotoSketchedRealitySketching2022}.
                This relationship has also been echoed in the context of personal fabrication~\cite{stemasovRoadUbiquitousPersonal2021}, which highlights the conceptual similarity of ``design-for-XR'' and ``design-for-fabrication''~\cite{doganFabricateItRender2022}.
                Nebeling further highlights the need for collaboration in content creation~\cite{nebelingXRToolsWhere2022}, which, in the context of architectural design~\cite{carpoSecondDigitalTurn2017} and personal fabrication~\cite{flathCopyTransformCombine2017} may be mediated through the exchange and remixing of parametric designs~\cite{oehlbergPatternsPhysicalDesign2015}.
                Customizer interfaces are a way to create remixes~\cite{flathCopyTransformCombine2017}, which was considered as a way to support creative practitioners in authoring immersive experiences through \emph{sampling}~\cite{stemasovImmersiveSamplingExploring2023}.
                Specific tools like SpatialProto~\cite{mullerSpatialProtoExploringRealWorld2021}, ProcessAR~\cite{chidambaramProcessARAugmentedRealitybased2021}, ProtoAR~\cite{nebelingProtoARRapidPhysicalDigital2018}, VRCeption~\cite{gruenefeldVRceptionRapidPrototyping2022}, or VRFromX~\cite{ipsitaVRFromXScannedReality2021} are examples where XR is leveraged to provide an authoring environment for new, immersive environments and applications.
                
                With our work, we do not target digital environments for the designs users may make, but rather physical ones, which demand transfers and understanding of requirements across physical (\emph{usage}) and digital (\emph{design}) environments.
                
Several works rely on path design~\cite{huttonpospickCreatingManipulating3D2023} or sketching as a way to define curves~\cite{baeEverybodyLovesSketch3DSketching2009,baeILoveSketchAsnaturalaspossibleSketching2008} or express design intent~\cite{igarashiTeddySketchingInterface1999,cambaSketchBasedModelingMechanical2022}, which is also a relevant parameter type in \system.
                This was also considered from an evaluation perspective by surveying the state of the art of how sketch-based systems are evaluated and highlighting the absence of standardized methods to do so~\cite{machucaMoreComprehensiveEvaluations2023}.
                Barrera Machuca et al. further outline the \emph{types of sketches} and \emph{interactions} used in systems.
                \system's curve drawing functionality primarily focuses on letting users generate ``conceptual sketches''~\cite{mayradonajibarreramachucaInteractionDevicesTechniques2023} to infer more precise geometry.
                Teddy by Igarashi et al. is considered a seminal interface for 3D design~\cite{igarashiTeddySketchingInterface1999} where a free-hand sketch is converted to a numerical representation without burdening the user with a detailed definition of it, going from rough expression to a finished design.
                
                \system deliberately provides ways to use coarse gestures to define curved elements instead of requiring users to precisely manipulate control points~\cite{igarashiTeddySketchingInterface1999} or deal with 3D-representations.

             \section{Concept and Design Rationale}\label{sec:prestudy}
    Before presenting the implementation of \system, we first outline the notion of \concept.
    
    \subsection[Novice-Facing Configurators and Parameters]{Novice-Facing Configurators \& Parameters}\label{sec:survey} 
        As a formative exploration, we initially examined how parametric designs are presented to (end-)users through the lens of configurator tools.
        Configurator interfaces enable ``mass-customization''~\cite{leeUnderstandingRolesIntelligent2022,nachtigallONEDAYShoesMaker2019} (i.e., highly personal designs at scale) and are \emph{interfaces to parametric designs}.
        In this work, we focus on users' \emph{interactions with parametric designs}, which are currently mediated through these \emph{configurator tools} or so-called \emph{customizers}.
        
        \paragraph{Commercial Configurator Interfaces} Tylko is an example of commercial furniture customization\fn{https://tylko.com/}{30.01.2024}.
        For a sideboard, users choose styles (which translate to entire geometric setups) and set dimensions (width, depth, height).
        The width parameter is \emph{continuous}, whereas height and depth are \emph{discretized} to 10 and 4 levels, respectively.
        Users can also alter colors based on a selection and choose whether they want back panels (boolean).
        To support users in estimating sizes, dimensions can be shown everywhere, a silhouette of a person is rendered next to the furniture, and users can toggle the display of household items (e.g., books, a laptop) in the respective depiction of the furniture.    
        The jewelry configurator ``Cell Cycle''\fn{https://n-e-r-v-o-u-s.com/cellCycle/}{30.01.2024} by Nervous System is a more sophisticated example: users not only define sizing and twist parameters, but can also alter the Voronoi-like pattern through direct manipulation to merge and subdivide cells or morph the entire geometry.
        As with the other configurators, there are options to choose a base type (e.g., cuff, ring), or a material, and a direct display of the price is shown. 
        In the space of personal fabrication, \tv offers a rich selection of customizable designs~\cite{liangCustomizARFacilitatingInteractive2022,oehlbergPatternsPhysicalDesign2015} that are based on openSCAD. 
        An example of such a design is ``Stretchy Bracelet'' by emmett\fn{https://www.thingiverse.com/thing:13505}{28.01.2024} from 2011, which is among the most commonly made designs on \tv.
        Noticeably, the parameters in that design are labeled as \texttt{r2}, \texttt{h}, \texttt{w}, \texttt{t}, \texttt{n}, and \texttt{m}, with some being integer numbers, and others floating-point.
        Nevertheless, the over 250 times it was fabricated by other users exhibit a variety of sizes and colors.
        A more recent yet similarly popular example is the ``Customizable drawer box'' by gpvillamil\fn{https://www.thingiverse.com/thing:421886}{28.01.2024}.
        Here, parameter names are more descriptive, like \texttt{drawer height}, and more diverse: users can enter text as a parameter (\texttt{message}) to be embossed or choose options between which object to generate (\texttt{part}).
        Here, parameters are grouped into categories like ``basic'', ``pattern'', and ``advanced'' to structure them further.
        A collection of vases by Ferjerez on \tv\fn{https://www.thingiverse.com/thing:2638924}{29.01.2024} consists of a single parametric design that allows users to choose from different styles (\texttt{shape}) that define a base shape, and a set of parameters to alter the vase's surface (e.g., \texttt{spikes}, \texttt{levels}).
        The author further provides a set of 50 already finished designs for users to fabricate directly to allow users to \emph{choose}, instead of customizing.\\
        
        From the examples above, various parameter types can be extracted: dimensional parameters related to extents, rotations, or positions, repetitions, choices from options, feature toggles, and other examples (e.g., media inputs like text or images).
        However, for most parameter types, there are underlying, more abstract classifications: functional parameters (e.g., related to press-fit), aesthetic parameters, body-related parameters, environment-related parameters, or parameters relevant to fabricating a design.
        We do not consider this list to be all-encompassing but rather an initial exploration of adopted customizer tools. 
An underlying issue is the configurators' ex-situ nature: they require users to design a bracelet without the arm it will be worn on, a shelf without the room it will be in, and without its future contents, thereby omitting the consideration of all \emph{physical interactions} between design and context.

    \subsection[Concept: In-Situ Interaction with Parametric Designs]{Concept: In-Situ Interaction with PDs}
        We consider the concept of \concept to be a way to enhance the expressivity of interfaces focused on retrieving artifacts (e.g.,~\cite{stemasovMixMatchOmitting2020,giunchiMixingModalities3D2021}), without demanding complex modeling tools.
        Leveraging extended reality as a vehicle to customize designs for personal fabrication may change how we treat \textbf{inputs} to such a customizer or parametric design. 
We argue that \concept may permit the creators of customizer systems to embed and expose more domain knowledge~\cite{baudischPersonalFabrication2017,leeUnderstandingRolesIntelligent2022} through the tools.
        This can happen through recommendations but also early evaluation, all within the users' unique physical contexts--knowledge often unavailable to authors od customizer tools. 
        Treating an XR headset as a sensor platform, then, allows \textbf{support} tailored to users' environments (e.g., by providing suggestions grounded in the device's scene understanding~\cite{stemasovShapeFindARExploringInSitu2022}).

        \paragraph{Target Audience} 
        With \system, we aim to support novices to design who may lack domain knowledge but still have unique requirements (e.g., how they want their spaces, for instance, at home, to look like).
        They may be able to express vague requirements~\cite{kochSemanticCollageEnrichingDigital2020} (e.g., labels~\cite{stemasovShapeFindARExploringInSitu2022}) to find a correct PD, and define specifics in-situ later on by exploring and iterating. 
        
        \paragraph{Target Expressivity and Objects} 
        We further focus on the configuration of everyday household artifacts (e.g., furniture) with functional and aesthetic components.
        They may not necessarily demand the degree of precision afforded by industry-grade design tools, but exhibit room for personalization to benefit end-users. 
        We further assume that they are either manufacturable with current or emergent means of personal fabrication or through providers capable of accommodating personalized artifacts (e.g., a print service).

    \subsection[Design Goals for pARam]{Design Rationale \& Design Goals for \system}
        Based on the previous considerations, assumptions, and prior art surrounding personal fabrication (e.g., \cite{liangCustomizARFacilitatingInteractive2022,stemasovShapeFindARExploringInSitu2022,leePosingActingInput2016}, we outline the following high-level design goals (DGs) for \system to embody the concept of \concept:\newline
        \textbullet~\textit{\textbf{DG1:}}\label{dg1}
            \textbf{\textit{Enable in-situ interaction with parametric designs}} for end-user configured, manufacturable artifact designs.
            To do so, we leverage extended reality as a vehicle to foster in-situ previewing, configuration, and evaluation. This is inspired by prior art in configuration tools~\cite{leePosingActingInput2016}, and, to a degree, research in measurement-to-design transfers~\cite{ramakersMeasurementPatternsUserOriented2023,kimUnderstandingUncertaintyMeasurement2017}. Most importantly, this tries to tackle the status quo of ex-situ interactions with parametric design configurators that are meant for consumers, as outlined in \autoref{sec:survey}, while also increasing the potential expressivity of search-based in-situ tools for personal fabrication (e.g.,~\cite{stemasovMixMatchOmitting2020}). 
\newline 
        \textbullet~\textit{\textbf{DG2:}}\label{dg2} 
            Support users in \textbf{\textit{choosing valid and fitting parameters}} with respect to their unique physical contexts and requirements. 
            While this is inherent to some customizer interfaces, they often focus on geometrical validity \cite{michelDAGAmendmentInverse2021} or fabricability~\cite{shugrinaFabFormsCustomizable2015} and less on the practicality and feasibility of a fabricated outcome. 
            Addressing this is enabled and supported by the use of extended reality as a previewing and sensor platform \cite{weichelMixFabMixedrealityEnvironment2014}. 
            This also involves supporting the transfer and conversion of context information to parameters, which may then reduce errors with respect to how a design interacts with the physical environment~\cite{ashbrookAugmentedFabricationCombining2016,mahapatraBarriersEndUserDesigners2019}.
\newline
        \textbullet~\textit{\textbf{DG3:}}\label{dg3}
            Support users in \textbf{\textit{understanding the effects and impact}} of their chosen parametrization early on and with respect to their physical environment (i.e., context).
            This is linked to \hyperref[dg2]{DG2}, and aims to provide an early understanding of the consequences~\cite{leighAfterMathVisualizingConsequences2015} of acquiring or fabricating a designed object.
            In the context of personal fabrication, the most fundamental ``consequence'' of a design or configuration activity is the physical acquisition of an object, which resides in a real and physical environment and \emph{interacts} with it \cite{stemasovMixMatchOmitting2020}. 
Context-specific support and understanding can, therefore, facilitate easier and more successful configuration processes by providing earlier previews without fabricated prototypes.\\

We relied on these design goals to guide the development of \system, but also argue that their underlying principles apply to fabrication-oriented design tools beyond parametric designs.
        
 \section{\system}\label{sec:system}
    The following sections outline the basic interaction flow for \system, followed by descriptions of the functionalities \system provides to enable and support the in-situ interaction with parametric designs.
    \begin{figure*}[h!]
            \centering
            \includegraphics[width=\minof{\linewidth}{0.9\textwidth}]{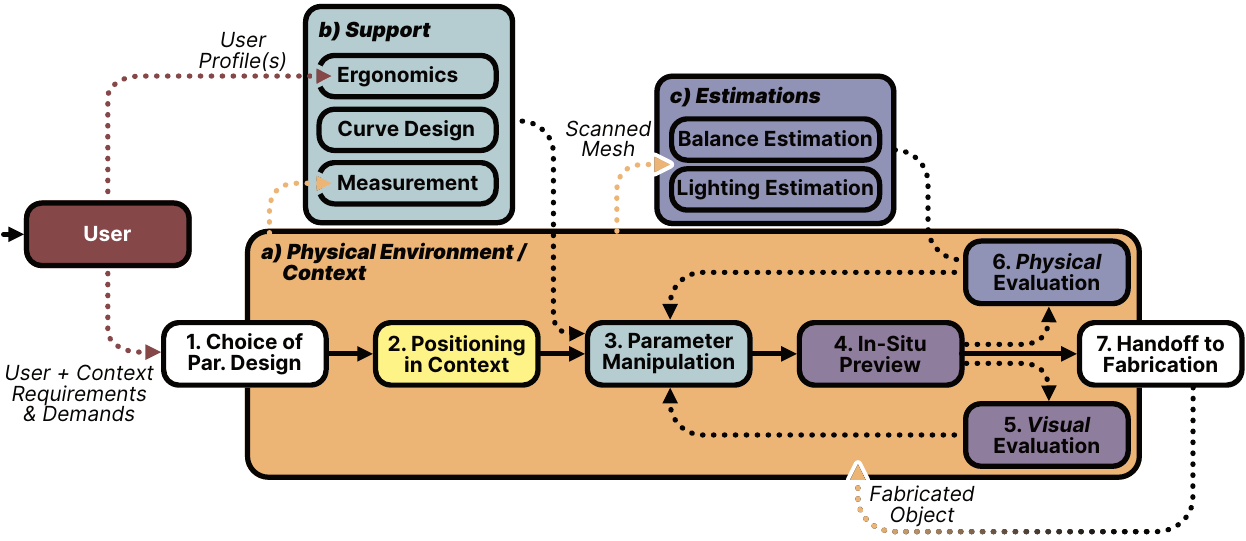}
            \caption{Overview of the process and functionalities supported by \system. \system enables in-situ interaction with parametric designs for personal fabrication (a). Relying on extended reality headsets facilitates in-situ previewing, but also input support like direct measurements (b) or estimation components that rely on the scanned environment mesh (c). Our implementation of \system focuses on steps 2 to 6 primarily, but covers the entire pipeline.}
            \label{fig:process}
        \end{figure*}

    \subsection{Envisioned Interaction Process}
        \autoref{fig:process} depicts a process of how users may customize parametric designs using \system.
        Users start by choosing an available design to customize. This choice is informed by their requirements (e.g., the desire for a new piece of furniture) and the context (i.e. where this furniture will be used later on).
        Both aspects are highly personal and, in their combination, may be unique to each user and situation~\cite{leeConsumerCreatorHow2017}.
        The user can then position the design at the location of future use (i.e., where it will be put after fabrication, \hyperref[dg1]{DG1}).
        For a shelf, for instance, this may be in an alcove or niche the shelf is supposed to fill.
The configuration process is supported (\autoref{fig:process}b) through embedded domain knowledge (i.e., ergonomic suggestions for body-related objects), in-situ measurements, or gesture-based input (e.g., for curve definition, \hyperref[dg2]{DG2}).
        The configured object can be evaluated visually. By leveraging the scanned environment mesh, estimations (\autoref{fig:process}.c) like the effects of a design on lighting and shadowing or the object's stability on the surface can be previewed and estimated early (\hyperref[dg3]{DG3}).
        After this validation step, users may choose to fabricate the object they configured, ultimately adding it to the physical context in which they configured it. 
        We consider this fabrication step to be beyond the scope of the paper and deliberately focus on context and user requirements in the design step of the personal fabrication pipeline.
        Notably, the process differs from parametric design tools established in personal fabrication (e.g., openSCAD) in two ways: 1) it situates all design activities in-situ, and 2) it facilitates in-situ iterations \emph{prior to fabrication} and aims to require no fabricated iterations.
    
    \subsection{Implementation}
        \system was implemented for the Microsoft HoloLens 2\fn{https://www.microsoft.com/en-us/hololens}{25.01.2024} using Unity 3D 2021.2.15.
        Two libraries were crucial for the implementation: MRTK 2.7.3\fn{https://github.com/microsoft/MixedRealityToolkit-Unity}{25.01.2024}, to handle inputs and interactions using the HoloLens, and Archimatix,\fn{https://www.archimatix.com/}{26.01.2024} to author custom parametric models for the application.
        Archimatix was found to be particularly suitable for development of \system, as it is tightly integrated with the game engine used, yet provides a rich and flexible environment to \emph{author} parametric designs.
        Alternative approaches like 1) using a networked instance of Grasshopper\fn{https://www.grasshopper3d.com/}{30.01.2024}, which would introduce latency and integration effort, or 2) pre-generating all possible variations of a design (e.g., using openSCAD), which would not scale, were considered, but ultimately discarded in favor of Archimatix.
        In total, \system provides 14 designs: 8 lampshades or vases, 2 tables, 3 shelf units, and a design for generic seating furniture (cf. \autoref{fig:models} in \hyperref[sec:appendix-a]{the appendix}).
        Additionally, \system also included a deliberately simple design (3 parameters only) for a book holder, used in the study (cf. \autoref{sec:evaluation}) for users to familiarize themselves with the tool.

        \begin{figure}[h!]
            \centering
            \includegraphics[width=\minof{\columnwidth}{0.65\textwidth}]{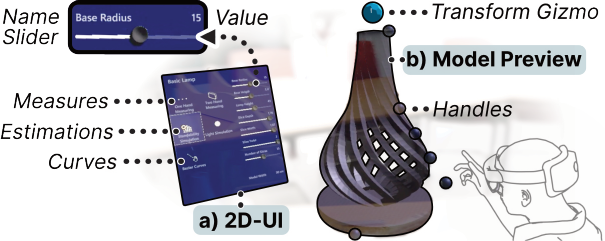}
            \caption{Basic elements of \system's user interface. A 2D UI (a) provides buttons to toggle all input and estimation modes and displays all parameters with a name, a slider, and a value display. A 3D model preview (b) allows the user to position the object in space and manipulate handles, depicted as blue spheres, to control individual parameters directly.}
            \label{fig:ui-overview}
        \end{figure}
    
    \subsubsection{Fundamental Parameter Interactions}
        The following interactions are heavily inspired by the ``state of the art'' in product configurators and parametric design (\autoref{sec:survey}).
        They transfer interactions to a mixed-reality environment to enable in-situ configuration and previewing of designs.
        \system provides 3 fundamental interactions with the design: positioning, slider-based changes, and changes through direct manipulation (\hyperref[dg1]{DG1}).
        An overview is depicted in \autoref{fig:ui-overview}, with the following paragraphs elaborating on individual features and their implementation.

        \paragraph{Positioning and Rotation} 
Users can initially interact with the most basic parameters of a parametric design: the position and the orientation of the object (\autoref{fig:ui-overview}b), which, in ex-situ configurators is deliberately omitted.
This allows them to position the digital preview at the location where they want to have the fabricated result\footnote{see also \autoref{fig:positionAndRotation}, \hyperref[sec:appendix-a]{appendix}}.
This is enabled by a gizmo above the model, provided by the MRTK package. 
            Users can position it with three degrees of freedom simultaneously and rotate it along the y-axis (\autoref{fig:ui-overview}b).
            This constraint is appropriate for all objects provided through \system as they are objects with a single, appropriate up-direction. 
            Scaling is not part of this manipulation, as it is covered by the parameters of each model. 

        \paragraph{Sliders}
As established in most expert and consumer-facing tools, sliders can be used to alter numeric values.
The limits of the slider convey basic constraints and ranges for each parameter.
            The sliders are shown on a 2D user interface (UI), seen in \autoref{fig:ui-overview}a.
This allows users to control separate dimensions of a design, reduced to a linear mapping\footnote{see also \autoref{fig:uiManip}, \hyperref[sec:appendix-a]{appendix}}.
The effects of the changes are reflected immediately in the positioned design.
            Users can position the UI anywhere in the space (world-fixed), but can also recall it to their hands immediately, by looking at their palm.
            The interface is then attached to the respective hand and can be positioned in the space again.

\begin{figure}[h!]
                \centering
                \includegraphics[width=\minof{\columnwidth}{0.65\textwidth}]{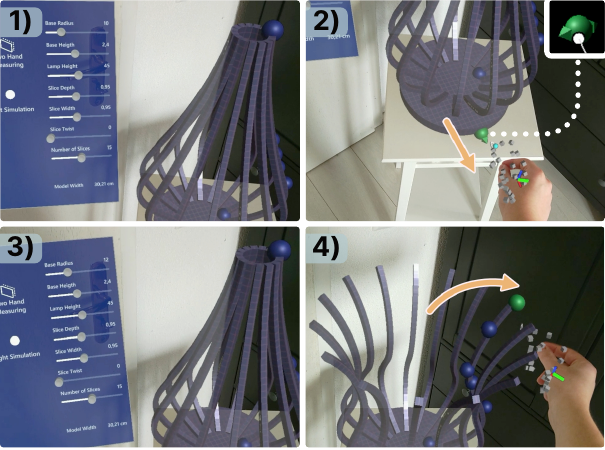}
                \caption{As an addition to slider-based interactions, \system lets users directly manipulate aspects of the model that are described by parameters, as an alternative to the 2D UI (1). Users can grab the spheres to change associated parameters (2). A direction hint is shown for parameters linked to an axis (2, top right). After letting go, the respective parameter is updated in the UI (3). Multidimensional parameters (e.g., positions), can be altered in a similar fashion (4). }
                \label{fig:directManip}
            \end{figure}

        \paragraph{Direct Manipulation}
Users can also use direct manipulation to interact with the parameters using handles (\autoref{fig:directManip}).
This conveys the relationship between a parameter and its effects on geometry more directly and avoids mediated interaction through a 2D UI~\cite{michelDAGAmendmentInverse2021,mcelroyPotScriptVisualGrammar2023}.
To manipulate parameters through the model, users can pinch any of the spheres positioned at and around a model and start moving it.
            The model is updated accordingly and at interactive speeds.
            Where fitting, the spheres also show arrows representing an axis on which the parameter can be altered, as an additional support mechanism.
            This can be seen in \autoref{fig:directManip}.2, where a radius is altered, and the arrows point along the radial direction.

\begin{figure}[h!]
                \centering
                \includegraphics[width=\minof{\columnwidth}{0.65\textwidth}]{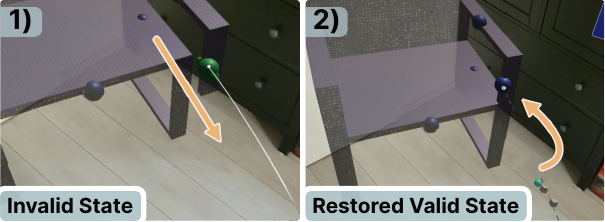}
                \caption{For parameter manipulations, \system allows users to temporarily ignore validity constraints (1), where the armrest protrudes beyond the seat, but restores a valid state of the model as soon as the user ends their interaction (2), with the model visibly snapping back to a valid state.}
                \label{fig:constraints}
            \end{figure}

        \paragraph{Constraints}
In addition to the range constraints for each parameter being indicated in the UI (i.e., the lowest and highest values of each slider), \system also includes interconnected constraints.
            Interconnected constraints describe valid ranges of specific sliders in relation to others. 
            For instance, a chair's armrest generally should not extend beyond the chair's seating surface to ensure connected geometry and, therefore, a degree of stability (\autoref{fig:constraints}.1).
Constraints like these ensure valid objects, as encoded by the person creating the parametric designs.
Whenever a user alters a parameter in an invalid fashion, the model ``snaps back'' to a valid prior state (\autoref{fig:constraints}.2).
            Constraints are set by the PD's author and can be absolute (e.g., a maximal value) or relative (e.g., a maximal value that changes depending on a different parameter value).
            In \autoref{fig:constraints}, the maximum extent of the armrest depends on the depth of the seating area, making it a relative constraint.
This further supports the notion of parametric designs, exclusively producing valid outcomes if the design's author sufficiently constrained the output space~\cite{shugrinaFabFormsCustomizable2015}.
        
    \subsubsection{Input Support}
        The second group of functions relates to \hyperref[dg1]{DG1 (in-situ interaction)} and \hyperref[dg2]{DG2 (support for parameter choice)}.
        \system leverages the capabilities of a mixed reality headset to support users in previewing designs, but also in configuring them.

\begin{figure}[h!]
                \centering
                \includegraphics[width=\minof{\columnwidth}{0.65\textwidth}]{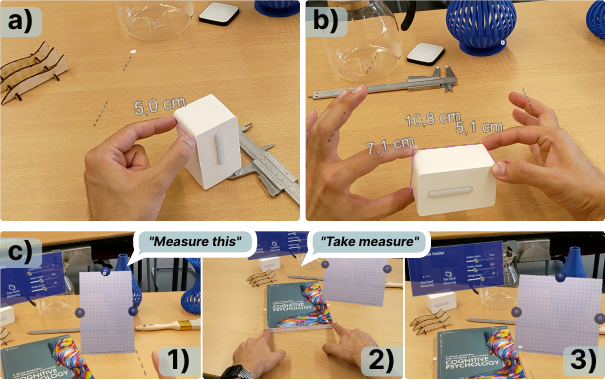}
                \caption{\system provides 2 modes for gesture-based measuring: one-handed (a), where the distance between index finger and thumb is used, and two-handed (b), where the distance between index fingers is taken. This feature can be used as a reference, or as a direct input to a parameter chosen previously. To leave the hands free for such measurements, \system provides voice interactions (c). Users can first point at a parameter to change, say ``measure this'' (1), move their fingers to the desired distance, and say ``take measure'' (2) to transfer the current value to the previously chosen parameter (3)}
                \label{fig:measure}
            \end{figure}

        \paragraph{Gesture-Based Measurement}
\system provides a simplified way to measure distances through gestures, inspired by prior art in HCI~\cite{leePosingActingInput2016,leeInteractiveSituatedGuidelines2018}.
Given that most parameters relate to measurements and dimensions, this aspect is crucial to consider.
            Error-prone transfers between analog measurement devices and digital design environments can either be bridged through connected measurement devices (e.g.,~\cite{weichelSPATASpatioTangibleTools2015}) or other support approaches~\cite{yungPrinty3DInsituTangible2018}.
Users can generate inputs to parameters through gesture-based measurements in two ways: one-hand measurements and two-hand measurements. 
One-handed measurements are taken between the user's thumb and index finger on one hand (\autoref{fig:measure}a).
Two-handed measurements are taken between the user's index fingers (\autoref{fig:measure}b).
            Both gestures are reasonably intuitive and are used in human-to-human communication of sizes and extents~\cite{koppSpatialSpecificityIconic2005,leePosingActingInput2016}.
Given that this feature may require both hands to operate, we settled on voice input to capture the measurements fluidly (\autoref{fig:measure}c).
            Users initially point at a slider (on the UI) or handle (on the model) and say \textit{``measure this''} (\autoref{fig:measure}c-1), which brings up a ruler between their fingers.
            They can then measure relevant interacting objects (e.g., the circumference of a plant while designing a pot).
            When they have settled on a length, they may say \textit{``take measure''} to record the value (\autoref{fig:measure}c-2) and directly alter the chosen parameter (\autoref{fig:measure}c-3).
            Users may also use this functionality to get an impression of size values in isolation or with respect to the context.
            The feature is based on the open-source work of Hiromu Kato\fn{https://github.com/HiromuKato/MRTK_HKSample}{27.01.2024}.
            We observed a measurement error between approximately 0.2 and 1cm across the tracking range.
            This was done informally by 2 of the authors repeatedly testing both measurement gesture modes along a ruler or a tape measure, but is roughly in line with results in literature~\cite{soaresAccuracyRepeatabilityTests2021,guoHoloLensTechnicalEvaluation2022}. While it may not achieve impeccable precision, this interaction is likely valuable~\cite{leeInteractiveSituatedGuidelines2018,leePosingActingInput2016} and will become even more useful given better tracking mechanisms. The measurement support allows users to measure interacting objects in-situ and immediately transfer these measurements to their design.
            This additionally allows users to take more complex measures, for instance, around irregularly shaped objects without employing calipers.

\begin{figure}[h!]
                \centering
                \includegraphics[width=\minof{\columnwidth}{0.65\textwidth}]{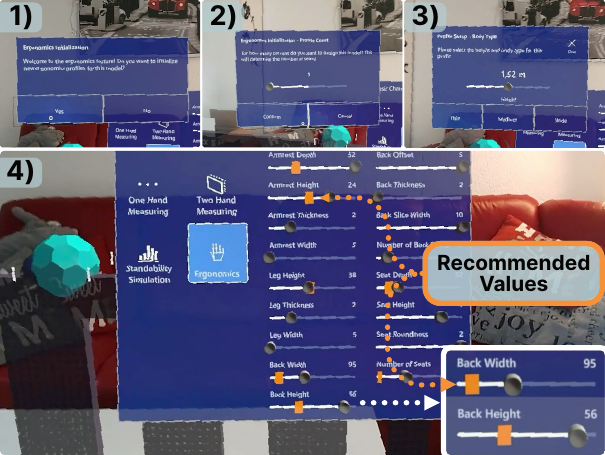}
                \caption{To support the configuration of body-related objects (e.g., furniture), \system allows users to configure a profile (1), which consists of a number of users (2), and their individual physical composition (3). Based on these profiles, recommendations for parameter ranges are shown on the slider interface in orange (4). The ranges provide a viable starting point, and can be based on one or several such profiles, reconciling parameter ranges across a set of differently-sized future users.}
                \label{fig:ergonomics}
            \end{figure}

        \paragraph{Ergonomics Recommendations}
\system also aims to simplify the way inputs are made to designs related to users' bodies (i.e., ergonomics-related ones).
            Home artifacts interact with users' bodies differently across different dimensions (e.g., prolonged/briefly, different body parts), which has been woven into expert-oriented~\cite{leeSmartManikinVirtualHumans2019,liliParametricDesignOffice2010} and novice-friendly~\cite{saulSketchChairAllinoneChair2011} design tools before.
            For seating-related objects (for \system, tables, and chairs or benches), \system provides a way to configure a set of users and their physical properties to receive recommended parameter ranges, where applicable.
While constraining the parameter space is inherent to parametric designs, they often lack recommendations as to how to navigate this constrained input space to achieve desirable results.
            Dealing with aspects like ergonomics demands domain knowledge and often boils down to designing for the majority~\cite{eastmankodakcompanyKodakErgonomicDesign2011}.
            A balance has to be struck between expressivity (i.e., being able to generate a \emph{valid} design) and fit (i.e., arriving at a configuration that fulfills criteria not present in the design environment).
Users first toggle the creation of profiles for each future user (\autoref{fig:ergonomics}.1-3).
            A bench to be used by a family would then require a profile for each of the members.
            For each profile, users input aspects like height or body composition.
            \system then recommended ranges in the UI (\autoref{fig:ergonomics}.4), similarly to prior approaches~\cite{leeInteractiveSituatedGuidelines2018}.
            The ranges are based on previous works~\cite{jocherRaumpilot2010,openshawErgonomicsDesignReference2006,eastmankodakcompanyKodakErgonomicDesign2011}, in our case primarily on the guide by Jocher and Loch~\cite{jocherRaumpilot2010}.
Ergonomics is a highly complex field -- through this function, parts of the domain knowledge needed to make reasonable decisions are provided to users through the interface without over-constraining the design.

\begin{figure}[h!]
                \centering
                \includegraphics[width=\minof{\columnwidth}{0.65\textwidth}]{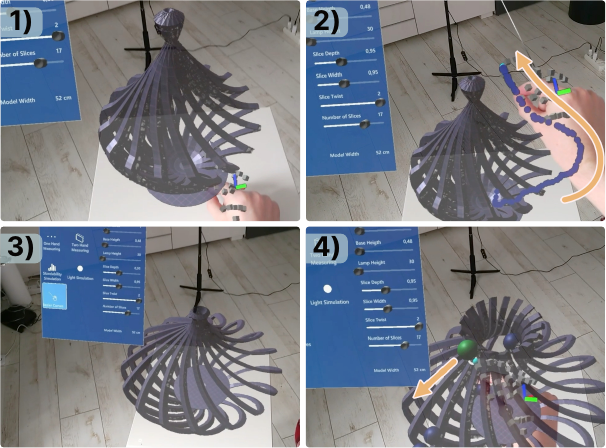}
                \caption{\system allows users to define curvature-related parameters using mid-air gestures. Starting from a base design (1), users can draw their desired silhouette in mid-air. Their input is visualized with a blue trace (2) and is converted to a B??zier curve and applied to the model (3). After this action, users can use this as a starting point for further parameter-based manipulations (4).}
                \label{fig:curves}
            \end{figure}

        \paragraph{Curves}
To simplify the definition of objects with curvature-related parameters, \system allows users to sketch their desired silhouette in mid-air (\autoref{fig:curves}).
Curves are generally complex to define, and parameters related to B??zier Curves (consisting of points and tangents) lose their spatial nature when treated as numerical parameters~\cite{yinNewBezierCurves2023}. 
            Prior works have deliberately obscured this representation~\cite{igarashiTeddySketchingInterface1999}, or focused on avoiding self-intersecting results~\cite{yukselClassC2Interpolating2020}. 
For objects consisting of curvature-related parameters, users can draw B??zier curves in mid-air to define how curved objects may look like.
            When presented with a design that supports such curvature parameters (e.g., lampshades or vases in \system), users can toggle a mode that allows them to draw their desired curvature (\autoref{fig:curves}.1).
            A pinch gesture lets \system track the motion, visualized by sampled points on the path (\autoref{fig:curves}.2).
            The path points are projected into a plane orthogonal to the user's view direction to convert the curve into a format suitable for Archimatix (i.e., one that operations like repeaters require).
            This 2D path is automatically converted to one or multiple cubic B??zier curves, potentially omitting details from the path, but also ensuring a smooth--``beautified''~\cite{mayradonajibarreramachucaInteractionDevicesTechniques2023}--result (\autoref{fig:curves}.3).
            Paths that ignore constraints (e.g., a minimal size) or are impossible to approximate with the design's number of B??zier curves are possible to draw, but are converted to a curve set that fulfills all constraints.
            Therefore, the result may differ from the sketched curve but remains valid.
            To correct the result inferred from the sketch, users can either re-draw the curve or correct it using the handles on the model itself (\autoref{fig:curves}.4).
            
    \subsubsection{Validation Support}
        Lastly, \system aims to provide a set of features to support validating designs with respect to the context they will be fabricated for.
        \system leverages the means of extended reality to support these context-dependent validations: the scanned environment mesh is an input for both validation components, the light estimation and the stability estimation.
        These estimation and validation components are not meant to be impeccable, industry-grade simulations (which are complex to set up and use), but rather quick and low-effort ways to understand the effects of design decisions \emph{in-context} (\hyperref[dg3]{DG3}).
        \system provides 2 sample components for that: a stability validation and a validation of lighting and shadowing patterns, elaborated in the next paragraphs.

\begin{figure}[h!]
                \centering
                \includegraphics[width=\minof{\columnwidth}{0.65\textwidth}]{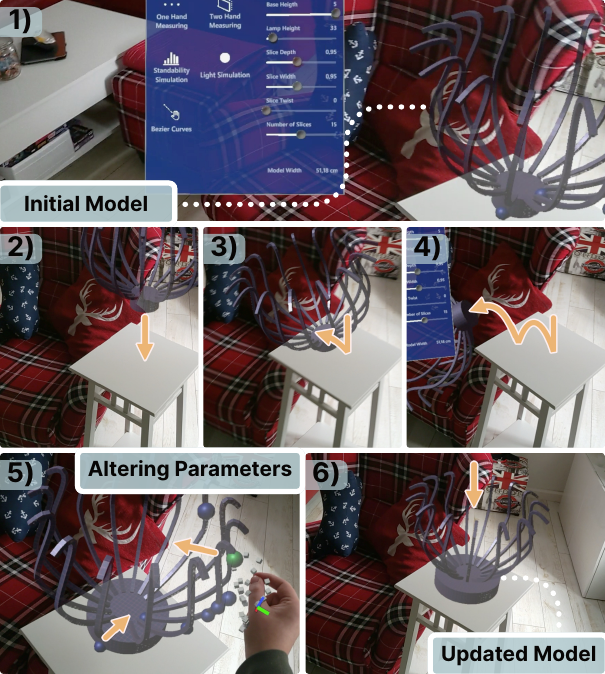}
                \caption{To understand the steadiness of designs, \system simulates a small drop of the model onto the scanned environment mesh. Users first position their model above its future position (1). It is then dropped onto this surface (2-3). A toppling design may still be functional but is indicative of an unsteady object. Users can then alter parameters, like widening an object's base (5), and re-try the estimation (6).}
                \label{fig:physics}
            \end{figure}

        \paragraph{Stability Estimation}
To evaluate the balance of an object, \system provides a rudimentary validation component for that (\autoref{fig:physics}.1).
            The configured model is dropped from a small height onto the environment (\autoref{fig:physics}.2-4).
            Stable designs do not topple from this motion, while top-heavy ones or designs with a small base surface do.
This allows users to get a glimpse of a design's stability and balance and react accordingly (\autoref{fig:physics}.5-6).
The feature relies on the scanned environment mesh provided by the HoloLens and the plane detection mechanism.
            The collision between the model and the surface is handled through the physics engine of Unity and uses default ``physics materials'' (i.e., configurations of mass and friction) and convex colliders for the model.
While this may not be a precise simulation (both due to the ``physics materials'' and colliders used), it can serve as a lower bound of stability regardless: a design that proves to be unstable at this point is likely to be unstable when fabricated.

\begin{figure}[h!]
                \centering
                \includegraphics[width=\minof{\columnwidth}{0.65\textwidth}]{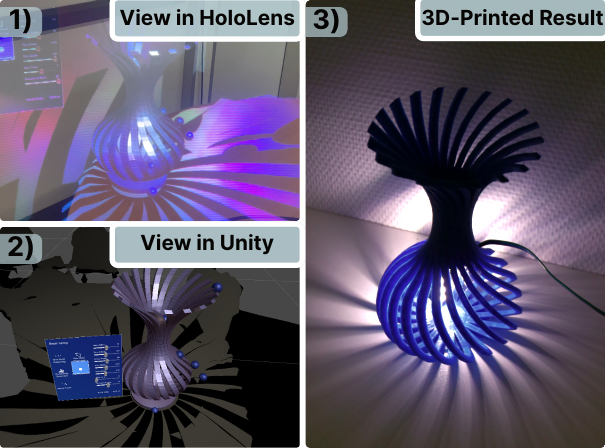}
                \caption{\system provides a function to estimate lighting and shadowing. This is relevant when designing lampshades and comparable objects. The shadowing pattern is projected onto the environments (1). Note that the effects of the shader are only visible on the device. This required a through-the-lens photograph instead of a screenshot for (1). Game engine view (2) and fabricated result (3) for reference.}
                \label{fig:lighting}
            \end{figure}

        \paragraph{Lighting and Shadowing Estimation}
\system also allows users to preview the effects a design may locally have on lighting and shadow patterns.
            This is particularly relevant for objects like lampshades, but can also be used with any object that interferes with light (e.g., new furniture obstructing existing light fixtures).
Lighting is a crucial design element in interior design contexts~\cite{karlenLightingDesignBasics2017, renDatadrivenDigitalLighting2023}, and any newly fabricated object is likely to alter this element at least to some degree.
            Furthermore, lighting is often non-trivial to estimate early~\cite{karlenLightingDesignBasics2017}, yet is a design element and component that is both functional (i.e., lighting has to be sufficient to read, for instance) and aesthetic (i.e., it has to be visually pleasing and potentially intriguing)~\cite{karlenLightingDesignBasics2017}.
Similarly to the stability validation, this feature likewise relies on the scanned environment mesh (\autoref{fig:lighting}.1). 
            A custom shader is applied to the environment mesh to display the shadow and light patterns (\autoref{fig:lighting}.2).
            This preview is based on a point light but could be expanded to other types of lighting if these relationships are encoded in the parametric design itself.
            Rendering shadows on an AR headset such as the HoloLens is non-trivial and requires bright light patterns and an off-black display of the actual shadows.
            As with the aforementioned stability estimation, this feature does not yield a perfect representation but rather an estimation (\autoref{fig:lighting}.3).
            Nevertheless, this estimation provides early, in-situ feedback on how a design works in a context when lighting is involved.

     \section{Evaluation}\label{sec:evaluation}\label{subsec:study}
We chose to evaluate \system and the underlying concept of in-situ interactions with parametric designs through a user study.

            \begin{figure}[h!]
                \centering
                \includegraphics[width=\minof{\columnwidth}{0.65\textwidth}]{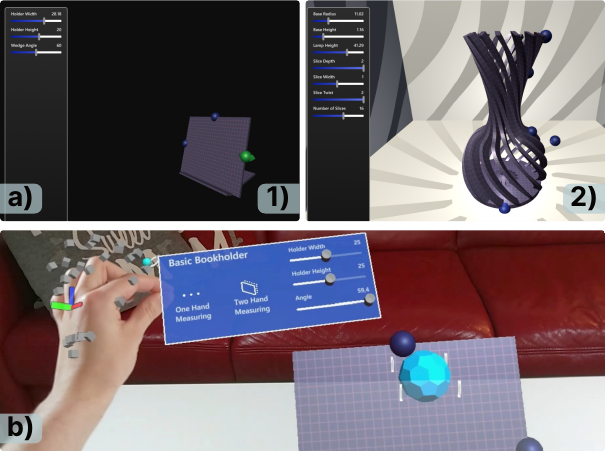}
                \caption{The user study compared \system with an analogous desktop version (a), which was similarly enriched with estimation capabilities (a2) and implemented following design patterns present in consumer-grade tools like openSCAD or the \tv customizer. A practice task, a bookholder with only 3 parameters was used for participants to familiarize themselves with the respective tools (a1--desktop, b--\system)}
                \label{fig:desktopVsAr}
            \end{figure}

            We chose to compare \system to a desktop-bound baseline-like application.
            This was meant to allow us to further understand the underlying concept of in-situ interaction with parametric designs, compared to what one may consider the status quo: ex-situ interaction to find and tune parameters.
            The underlying goal was to understand whether \system, as a proof-of-concept implementation, can transfer the benefits of parametric designs to XR while providing meaningful support through the physical context.
            A core goal in developing this custom baseline was to achieve feature-parity between desktop and AR versions, meaning that added functions (e.g., a light estimation) should be present in both versions.
            Nevertheless, the design of the desktop version followed patterns and paradigms present in user-facing parametric designs, such as openSCAD, the \tv customizer, commercial services like Tylko, or academic works like Fab Forms~\cite{shugrinaFabFormsCustomizable2015}.
            Specifically, they all follow an approach combining sliders, text fields, dropdowns, and a preview, occasionally used for direct manipulation.

        \subsection{Sample}
            The sample for the evaluation consisted of 21 participants.
            One of them had to drop out due to intermittent technical issues, leaving us with a sample of $n = 20$.
            6 of the participants identified as female, one person identified as non-binary, and 13 participants identified as male. 
            The sample had a mean age of 24.65 ($SD = 2.11$), with their backgrounds being mostly students (e.g., in finance, psychology, computer science).
11 participants had experience with VR and AR, specifically using headsets like the Meta Quest or the HoloLens.
            Out of these 11, 6 reported a prior usage duration of 1-10 hours, 3 reported 10-100 hours, and 2 reported over 100 hours of use.
            13 participants reported having used product configurators before (e.g., for cars, furniture, or clothing, specifically footwear).
            Similarly, 13 participants reported having used 3D-modeling tools before (e.g., AutoCAD, Blender, Tinkercad, SketchUp, but also design tools like Unity3D), which cover ``upper'' classes of content creation tools in the classification of Nebeling and Speicher~\cite{nebelingTroubleAugmentedReality2018}.
            Out of these 13, 3 reported a prior usage duration of 1-10 hours, 5 reported 10-100 hours, and 5 reported over 100 hours of use.
            We classify the 10 participants with over 10 hours of prior use as ``experts'' in the following steps.

        \subsection{Procedure}
            The study was conducted in a meeting room of our institution.
            Various pieces of furniture were present and were actively embedded in the task design (cf. \autoref{par:tasks}).
            The participants were recruited around our institution with the help of mailing lists.
            After a greeting and a brief introduction, participants signed consent forms.
            They then received an introduction to the respective tool they will be using (\system or its desktop counterpart).
            A simple model consisting of 3 parameters -- a book-holder -- was then used as a training environment for the participants to familiarize themselves with the tool (\autoref{fig:desktopVsAr}).
            Participants then received instructions regarding the task they would be solving.
            This information was also available as a printout for them to reference on demand.
            After finishing the first task, users filled out a post-task questionnaire.
            This was repeated for the second tool and concluded with a final questionnaire comparing the tools and acquiring demographic data.
The study took 60 to 70 minutes.
            Participants received a reimbursement of 12\texteuro~for their time.

            \paragraph{Tasks}\label{par:tasks}
                Participants were asked to solve 2 distinct tasks, with either \system or the desktop variant.
                The order and the tool-task combinations were counterbalanced using a Latin square.
                The two tasks were the configuration of a bench and the configuration of a lampshade. 
                Each task consisted of 3 main instructions and a bonus instruction.
For the \textbf{bench}, participants were asked to design a bench for themselves and another fictional person to use comfortably. 
                The other person was described as having an average physique and a height of 1.77 meters.
                Participants had to \hypertarget{rec:b1}{B1)}\label{rec:b1} design it in a way that it is as comfortable as possible for them and the fictional partner, \hypertarget{rec:b2}{B2)}\label{rec:b2} the width of the bench should not exceed the length of the table in the study room, \hypertarget{rec:b3}{B3)}\label{rec:b3} the armrest height should be aligned with the height of the side table in the study room,
                and optionally, \hypertarget{rec:b4}{B4)}\label{rec:b4} the bench should fit 2 seat cushions provided by the study coordinator.
For the \textbf{lampshade} task, users were asked to design a lampshade that is visually pleasing in terms of shape and lighting/shadowing patterns.
                The lampshade \hypertarget{rec:l1}{L1)}\label{rec:l1} should have a maximum height of 40cm, \hypertarget{rec:l2}{L2)}\label{rec:l2} its diameter must not exceed the width of the bedside table in the study room, \hypertarget{rec:l3}{L3)}\label{rec:l3} be stable and not topple easily, and optionally, \hypertarget{rec:l4}{L4)}\label{rec:l4} a provided candle should fit inside of the lampshade.
To solve the aforementioned tasks, participants had a set of measurement tools freely available to them: calipers, a folding rule, a small ruler, and a tape measure.
                Participants were also allowed to use them while designing with \system. All reference objects (e.g., the ``bedside table'') were easily accessible in the room and could be measured by the participants.

            \paragraph{Measures}
                For the user study, we acquired the following measures: the System Usability Scale~\cite{brookeSUSAQuickDirty1996}, the (raw) NASA Task Load Index (TLX)~\cite{hartDevelopmentNASATLXTask1988}, and the Creativity Support Index (CSI)~\cite{cherryQuantifyingCreativitySupport2014}.
                We further acquired a set of single-item Likert-scaled questions on 5-point scales and solicited usefulness rankings for each of the provided features (cf. \autoref{sec:system}).
                Participants were also asked to express modality preferences (i.e., AR vs. Desktop) for specific design tasks. 
                Successes and failures concerning individual tasks were acquired following the criteria outlined in \autoref{par:tasks}.
                Lastly, all interactions (i.e., features being used), were logged to allow for an analysis of patterns and strategies participants demonstrated.

        \subsection{Results}
            The following paragraphs outline the results we were able to gather from the evaluation.

            \begin{figure*}[h!]
                \centering
                \includegraphics[width=\linewidth]{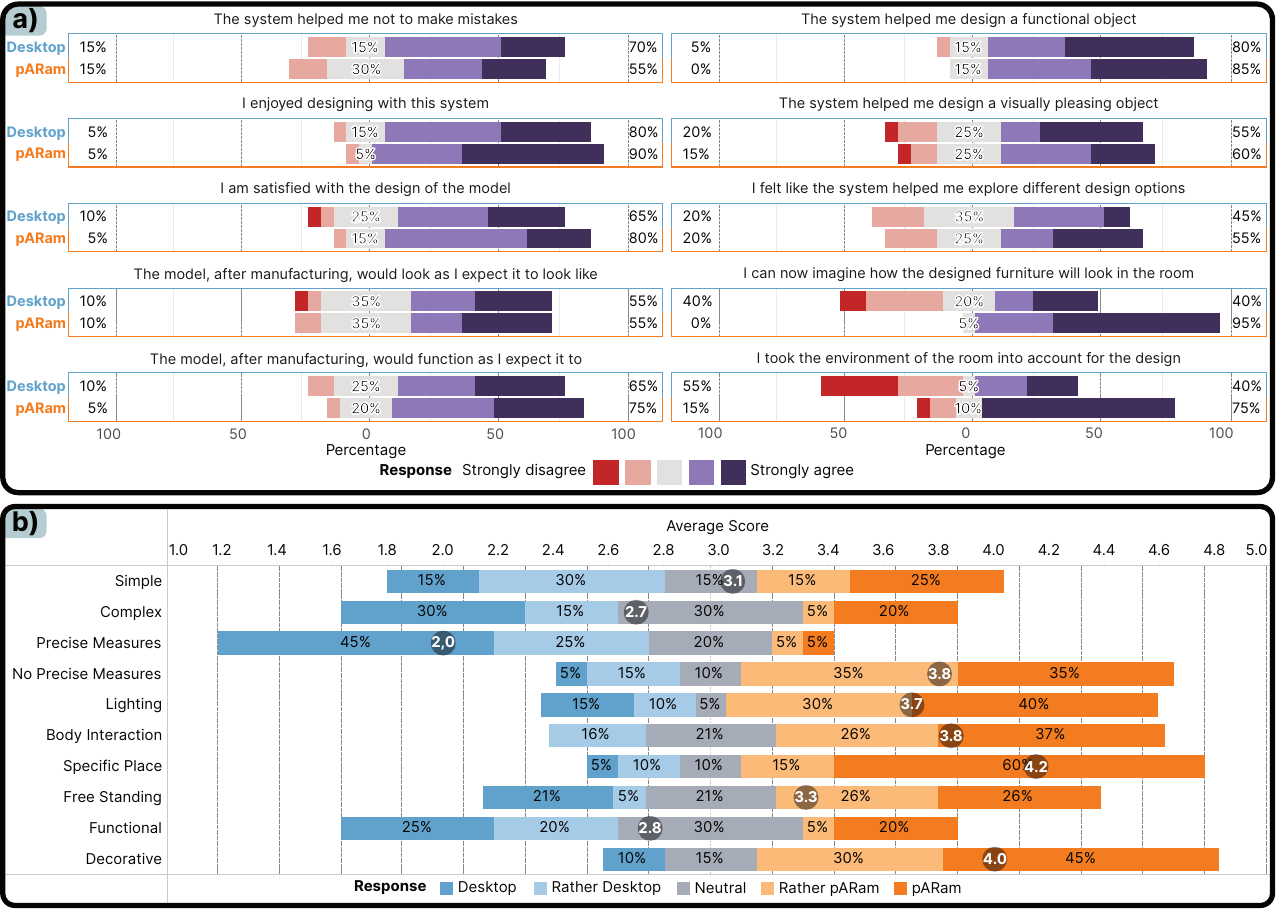}
                \caption{Overview over statement ratings and application preferences from the user study: a) Participants' stances towards the tool they used and towards the outcome of their design process using the tool, rated on a 5-point Likert scale; b) preferences towards applications (desktop and \system) rated by participants on a 5-point scale for different categories of objects.}
                \label{fig:plots}
            \end{figure*}

            \paragraph{Questionnaire Data}\label{par:questionnairedata} 
                For the analysis of the SUS and CSI scores, as well as for the individual TLX items, we apply the Aligned Rank Transform (ART) for nonparametric factorial analyses~\cite{wobbrockAlignedRankTransform2011}.
The ART found a significant main effect of the tool used on System Usability Score (\F{1}{18}{15.50}, \pminor{0.001}) (AR: \m{46.58}, \sd{11.38}, Desktop: \m{57.38}, \sd{7.98}).
                This can be traced back, at least in part, to suboptimal interactions with the spatial interface, especially the sliders on the 2D interface.
                This criticism was voiced during the study itself and was also found in the comment fields in the questionnaire.
The overall (raw) TLX score exhibited expected differences (AR: \m{3.56}, \sd{0.90})(Desktop: \m{3.00}, \sd{0.73}).
                Our analysis focuses on individual subscales.
                On the TLX subscales of Performance (AR: \m{5.60}, \sd{1.47} Desktop: \m{6.20}, \sd{0.95}), Frustration. (AR: \m{2.90}, \sd{1.62}, Desktop: \m{1.90}, \sd{0.85}), Mental Demand (AR: \m{3.60}, \sd{1.50}, Desktop: \m{3.05}, \sd{1.61}), and Temporal Demand (AR: \m{2.75}, \sd{1.71}, Desktop: \m{2.00}, \sd{1.12}) no significant effects were found. The ART found a significant interaction effect of the tool used $\times$ model type on TLX Effort (\F{1}{18}{11.73}, \p{0.003}) between AR (\m{3.65}, \sd{1.35}) and Desktop (\m{3.15}, \sd{1.46}).
                The ART further found a significant main effect of the tool used on Physical Demand (\F{1}{18}{9.68}, \p{0.006}) between  AR (\m{2.85}, \sd{1.57}) and Desktop (\m{1.70}, \sd{1.22}).
                In light of the inherently spatial and more embodied interaction fostered through the headset and the gestural interaction is a reasonable outcome.

For the Creativity Support index, the ART found no significant effects.
                The score for \system (\m{54.13}, \sd{15.98}) was comparable to the one for the desktop (\m{55.40}, \sd{13.94}).
                This was also the case when comparing ratings by experts (\m{55.08}, \sd{14.06}) compared to non-experts (\m{54.19}, \sd{16.66}).

For most single-item Likert-scaled questions (\autoref{fig:plots}a), no significant differences were found.
                However, the ART found a significant main effect of the tool used on the responses to \textit{``I can now imagine how the designed furniture will look in the room''} (\F{1}{18}{25.21}, \pminor{0.001}) between AR (\m{4.60}, \sd{0.60}) and Desktop (\m{3.15}, \sd{1.39}).
                The ART also found a significant main effect of the tool used on the question rating for \textit{``I took the environment of the room into account for the design''} (\F{1}{18}{24.46}, \pminor{0.001}) between AR (\m{4.30}, \sd{1.30}) and Desktop (\m{2.75}, \sd{1.59}).

            \paragraph{Task Successes and Error Causes} 
                Based on the tasks outlined earlier, we further conducted an analysis of successes.
Based on the main tasks given to each participant (\hyperref[rec:l1]{L1}-\hyperref[rec:l3]{L3} and \hyperref[rec:b1]{B1}-\hyperref[rec:b3]{B3}), 2 people failed (i.e., did not complete one or more of the tasks) the lamp task on the desktop and 3 did so using \system.
                For the bench task, 4 users failed at one or more of the mandatory subtasks using the desktop variant and 6 failed using \system.
When including failures in the optional tasks (i.e., \hyperref[rec:l4]{L4} and \hyperref[rec:b4]{B4}), 4 people failed the bench task using the desktop, and 6 failed with \system.
                For the lamp task, 4 people failed any subtask using the desktop variant, whereas 7 did so using \system.
                3 expert users made errors on the desktop, compared to 3 non-experts.
                With \system, 5 expert users made errors, compared to 4 non-experts.
When considering overall error extents, \system (\m{5.86}, \sd{5.33}) and the desktop counterpart (\m{5.33}, \sd{4.71}) exhibit comparable ranges. 
When considering individual tasks, the errors in the bench task on the desktop  (\m{6.48}, \sd{5.43}) are comparable to the bench task using \system (\m{6.5}, \sd{6.35}).
                The error extents for the lampshade task were lower, yet remained comparable between \system (\m{4.59}, \sd{3.02}) and desktop (\m{3,04}, \sd{2.63}).
All failures for the bench task stemmed from \hyperref[rec:b3]{B3} \preferr{4}{6}, a constraint regarding the armrest's height, where an error tolerance of 1cm in each direction was allowed.        The issues for the lampshade task were more spread out, with issues present in \hyperref[rec:l1]{L1 -- maximum height} \preferr{1}{1}, \hyperref[rec:l2]{L2 -- maximum diameter} \preferr{1}{2}, and \hyperref[rec:l4]{L4 -- fit of a candle (optional)} \preferr{2}{4}.

            \paragraph{Tool and Feature Preferences}\label{sec:prefs}
When asked to rate which tool they would rather use for a specific type of object on a 5-point scale, a set of trends became visible.
                They are depicted in \autoref{fig:plots}b.
                There were clear preferences for a desktop-based interaction for complex objects \pref{45}{25}, and objects that require precise measures \pref{70}{10}.
                No definitive trend was visible for simple objects \pref{45}{40} or functional objects \pref{45}{25}.
                In turn, there was a trend towards \system for objects that do not need precise measurements \pref{20}{70}, lighting-related ones \pref{25}{70}, objects that interact with the body \pref{16}{63}, decorative objects \pref{10}{75}, and, rather clearly, objects meant for a specific place \pref{15}{75}.
Users were also asked to rank the tool's individual features with respect to their helpfulness in solving the task given to them.
                For \system, the curve drawing functionality was rated as most helpful (\mrank{1.6}), followed by the handles (\mrank{2.22}), sliders (\mrank{2.65}), the measurement functions (\mrank{2.69}), the ergonomics recommendations (\mrank{3.25}), the stability estimation (\mrank{3.45}), and the light estimation (\mrank{4.2}).                    
                Comparable patterns emerged for the desktop version.
                Here, however, the sliders were generally ranked as most helpful (\mrank{1.6}), followed by the handles (\mrank{1.85), the ergonomics feature (\mrank{1.83}), the stability estimation (\mrank{2.79}), and the light estimation (\mrank{3.38}).
                It is intriguing to note that subjective helpfulness does not translate 1:1 to actual feature use, as seen in \autoref{fig:sequences}, given that fundamental interactions are likely indispensable but less impressive compared to more use-case-specific ones.
                \begin{figure*}
                    \includegraphics[width=\linewidth]{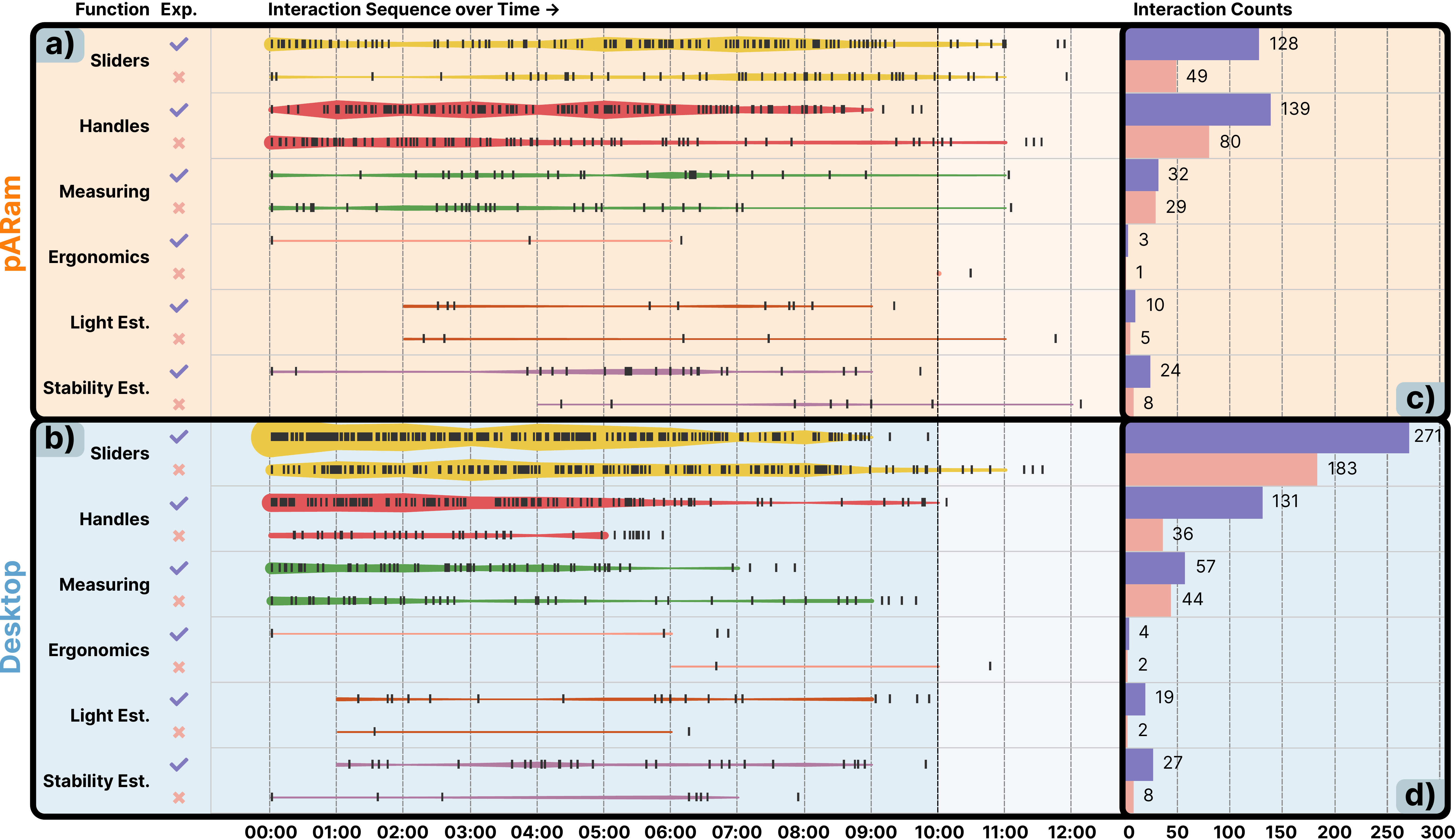}
                    \caption{Aggregated interaction sequences for \system (a) and its desktop (b) counterpart. The first column indicates the tool (desktop, \system), the second column refers to individual features, followed by a split between \textcolor{expert}{\textbf{experts}~(\checkmark)} and \textcolor{nonexpert}{\textbf{non-experts}~(\textit{x})}. The timeline in the fourth column depicts individual interactions, with their distribution further highlighted through a violin plot. The soft cutoff for the study is indicated with a dashed line at the 10-minute mark. The last column provides a count for each interaction in a line for \system (c) and desktop (d).}
                    \label{fig:sequences}
                \end{figure*}
            
            \paragraph{Interaction Sequences and Feature Use}
                All interactions with the tools were logged.
                An aggregation of interactions across all participants can be seen in \autoref{fig:sequences}.
                A complete set of per-participant timelines is found in \autoref{fig:interactionSequencePerUser} in \hyperref[sec:appendix-b]{the appendix}.
                Between the desktop version (\autoref{fig:sequences}b) and \system (\autoref{fig:sequences}a), a set of trends is further visible:
                Experts generally measured more and did so earlier in the process, and more measurement activities may also indicate task success.
                Measurement activities were also more scattered across the process in \system, likely due to their easy-to-access nature.
                In contrast, measurement often preceded most, if not all, activities for the desktop variant: users tried to acquire all relevant measures first and then alter their design under these constraints.
                The estimation functions of \system (light, balance) generally happened later in the process, given that they are meant to \emph{validate} finished designs or design iterations, but were used less often than expected.
                For both modalities, the estimation functions were used more commonly by experts, and using them may indicate success in related tasks (e.g., \hyperref[rec:l3]{L3}).
Interaction counts can be seen in \autoref{fig:sequences}c and  \autoref{fig:sequences}d. 
                Sliders were used more often on the desktop (\avg{23.9}) compared to the handle-based interactions (\avg{13.9}).
                In contrast, AR seemed to have fostered more direct interaction through the handles (\avg{12.2}), compared to the sliders on the 2D UI (\avg{9.3}).
                This is linked to both the highly spatial nature of the interaction and visualization, but also to the higher physical demand (cf. \autoref{par:questionnairedata}).
                Lastly, users were generally quicker to complete their tasks on the desktop.

            \paragraph{Observations and Open Comments} 
                We further solicited non-mandatory open comments about the positive and negative aspects of the respective tools.
In contrast to responses to standardized questionnaires like the SUS or NASA TLX, participants were highly enthusiastic about \system and the underlying concept.
                They stated that \system let them  \iquote{[...] iterate quickly and freely}{P10} (also voiced by P5, 15, 20), or \iquote{[...] let me imagine how it would look like at that exact place}{P6}. 
                This core principle underlying \concept was further appreciated with respect to \emph{validating designs}: \iquote{[...] I like the possibility to explore in a dynamic way how my decision affects the outcome of the design}{P15}, \iquote{It was quite easy to design a matching lamp because I could compare virtual and real world objects in place}{P21}, which was an \iquote{[...] an immense help}{P19}.
                It was further appreciated for being \iquote{[...] fun [...]}{P10}, which was also voiced by P12, 17, 20.
                The measurement support was also appreciated: \iquote{[...] the measuring tool made meeting the requirements really easy}{P19}, (in addition to P13, 20) and so were the estimation functions (P14, 21). 
                
As a downside, participants mentioned the basic input handling, especially related to the sliders on the 2D interface: the \iquote{Pinching gesture was very hard to do correct}{P3}, or \iquote{The use of the sliders was a bit difficult to get the exact value I wanted}{P14}. 
                Criticisms related to the HoloLens' inputs and MRTK's handling of them were also mentioned by most participants in some way (P1,3-5,8-12,14,16,18-21) and can be optimized through UI sizing, positioning, and sensitivity.
                The \iquote{Usable area of the glasses [was] very limited}{P3}, referring to the limited field of view of the HoloLens.
                This sentiment was echoed by other participants (P5, 6, 8, 17).
                
The desktop variant, in contrast, was praised for how well the interactions on a 2D screen worked with the sliders: it \iquote{very easy to design the model using the sliders}{P7}.
                Without in-situ references, recommendations became more relevant: \iquote{The ergonomics feature helped a lot to define required width and height of the furniture}{P21}.
For the desktop variant, participants voiced concerns regarding slider-parameter links more commonly (P4, 5, 19-21).
                Participants further suggested features for the desktop variant that borrow functions from \system: \iquote{[...] overlap [the model] with a picture of the room [...] to see how it looks like in an environment}{P15}. 
                Comparable sentiments were also present in other participants' comments (P1, 3, 6, 11).
                As with \system, participants also criticized aspects like view controls (P8, 9, 12, 16, 19) or discoverability (P10, 21).
                
                For most participants, we observed a vastly more spatial and dynamic approach to configuration with \system compared to the desktop variant.
                Not only did they design and preview in-situ, but they also fluidly moved between phases of measuring, designing, and evaluating (\autoref{fig:sequences}a).
                Using the desktop tool, measurement activities were more present earlier, and became rarer later on.
                An intriguing observation was that a small set of participants used the provided analog measurement tools with \system, prior to remembering its built-in measurement functionalities and fully switching to use them instead. \section{Discussion}\label{sec:discussion}
    The following sections engage with the study results and reflect on the concept of \concept.
    
    \subsection{Insights from the Evaluation}\label{sec:studydiscussion}
        From the user study, we gathered that \system, as a proof of concept, is not only functional and valid support for users, but also identified challenges underlying notions of low-effort--high-expressivity design tools.
        
        \paragraph{Spatial Interactions to Define Artifacts}
            The concept of \concept along with \system embraces the spatial nature of artifacts, but also the interactions to define them.
            Participants praised this nature: not only did they actively move around their design, but they also altered the position of the design freely, for instance, to make it easier to fulfill environment-related design tasks (e.g., \hyperref[rec:b2]{B2}, \hyperref[rec:b3]{B3}, \hyperref[rec:l2]{L2}).       
            An often-ignored parameter in various configurators (by their ex-situ nature) is an artifact's position in the user's (and, therefore, the designer's) context of use. 
            Embedding this parameter meaningfully in a design tool either requires full and impeccable acquisition of that context, or through situated interaction, as enabled by \system.
            Interacting through gestures is likely a highly approachable way to define objects \cite{holzDataMimingInferring2011,leePosingActingInput2016} and can be further enriched with modalities like speech \cite{ballagasExploringPervasiveMaking2019}.
            Spatial interactions, however, come at the expense of precision and fidelity, at least with the current state-of-the-art of tracking, which has to be considered upfront.
The tradeoff between achievable precision and spatial interaction is reflected in several questionnaire items, like the TLX score, individual items, or the SUS score. 
            It may require surface-supported interactions, as shown with DesignAR~\cite{reipschlagerDesignARImmersive3DModeling2019}.
            The comparable means in the CSI score may indicate that \system, as a creativity support tool, may be on par with the desktop counterpart, but this claim warrants further studies.
        
        \paragraph{Overconfidence through Gestural Measurements}
            The measurement precision and fidelity of users using \system was arguably below the one achieved with the desktop variant.
            Users were free to use the provided measurement tools, but most chose to rely on \system and gestural measurement. 
            This is likely grounded in the fact that transferring the acquired dimension \cite{weichelSPATASpatioTangibleTools2015} to the design tool (or even a specific parameter) was made deliberately frictionless.
            A user's fingers are, especially at small scale and low margins for error, no replacement for a caliper. 
However, with mechanisms to compensate for that (e.g., snapping~\cite{nuernbergerSnapToRealityAligningAugmented2016}, which takes context into account), they may approach a useful degree of precision.
While this is a technological limitation that can be rectified through better tracking and compensation algorithms, there is a tendency towards ``eyeballing'' or ``guesstimating'' when engaged in in-situ design.
            This notion of coarse measurements may be sufficient for isolated artifacts and will greatly reduce effort, but does not transfer to highly precise objects that functionally interact with their environment.
            Mahapatra et. al identified such issues (i.e., ``eyeballing'') in their work~\cite{mahapatraBarriersEndUserDesigners2019}, and we believe that direct visual feedback may reinforce those.
            Immediate, in-situ visualization of the future artifacts seems to create a tendency to operate on the visual preview, instead of precise and detailed measurements.
            Users may then treat this preview as the \emph{only} feature to evaluate a design, unaware of imprecisions in the model, the visualization, or their estimations.
            We consider this tendency towards ``eyeballing'' and subsequent errors to be the ``cost of the system'', which, in future iterations have to be corrected, for instance, by relating gestures to machine-measurable objects in the context~\cite{liangCustomizARFacilitatingInteractive2022} or snapping~\cite{nuernbergerSnapToRealityAligningAugmented2016}.
            Alternatively, approaches like the ones presented by Kim et al. \cite{kimUnderstandingUncertaintyMeasurement2017} or Ramakers et al. 
           ~\cite{ramakersMeasurementPatternsUserOriented2023} that allow for tolerant designs may also support imprecise, low-effort design processes while retaining functional outcomes.

        \paragraph{Correspondences between Interface and Model}
            In some of the designs in \system, there were a few parameters that were only controllable through handles, and some were only controllable through the sliders of the UI. 
            This led to perceptions of unintuitiveness, where participants were \iquote{[...] not completely sure what each of the blue spheres were doing}{P14}. 
            Subsequently, they engaged in exploring and discovering individual elements to understand their effects and constraints.
            Having both the model and the UI in view--which is non-trivial for users to do with the HoloLens--helped make logical connections, but highlighting and linking would likely be helpful in future iterations of \system.
           
    \subsection[Reflecting on In-Situ Interaction with Parametric Designs]{Reflecting on In-Situ Interaction with PDs}
        \paragraph{Interactions for Working Across Scales}
            With \system, both human-scale~\cite{kovacsHumanScalePersonalFabrication2021} or furniture-scale objects~\cite{agrawalProtopiperPhysicallySketching2015} (e.g., tables, shelves) and smaller trinkets (e.g., vases, lampshades) are customized through the same interaction techniques at different scales.
            While this helps, consistency, precision, and error tolerances are likely to change between these scales.
            This likely requires a more thorough consideration of gesture-based measurement, for instance, incorporating wider gestures, and combining pointing with other modalities \cite{leePosingActingInput2016}.
            XR devices further can provide immediate, visual, and in-situ feedback on such inputs.
            However, if these means for measurement are not given (or are not precise enough), users will likely revert to in-situ estimating and ``eyeballing'', which may work in some, but not all cases.
        
        \paragraph{Encoding and Evaluating Function Early} 
            Any fabricated artifact fulfills--ideally--user requirements in terms of form and function. 
            Users can choose to emphasize specific aspects of those, but they remain tightly coupled, given that they are designed for physical space, not only a digital one.
            Encoding function and physical validity \cite{shugrinaFabFormsCustomizable2015} is non-trivial.
            In the study, we noticed this with the optional candle task (\hyperref[rec:l4]{L4}), where users routinely emphasized form over this function, given that the artifact was one that is inherently associated with looks.
            Support for these tasks is crucial and is part of \concept.
            While \system attempted to provide this support, users had to actively choose to use and interpret it correctly (e.g., what to do when the stability estimation indicates a toppling design).
            For the ergonomics recommendations, this was not the case for all the participants (\autoref{fig:sequences}), nor was this the case for the light and stability estimations.
            This is an argument for automated validations, which can understand design intent in isolation \cite{hofmannGreaterSumIts2018}, but also in relation to the context.
        
        \paragraph{Bridging Physical Context and Digital Parameters}
            In-situ design allows for an overlap of usage context and design environment~\cite{yungPrinty3DInsituTangible2018,weichelMixFabMixedrealityEnvironment2014,stemasovMixMatchOmitting2020}, but does not free the user from engaging in some, potentially error-prone~\cite{mahapatraBarriersEndUserDesigners2019} transfers.
The measurement component in \system supports these transfers, but does not guarantee that the dimension or context object being measured by the user is correct, which was reflected in the error rates of \system.
            In turn, the previewing functionality enables the transfer from the design environment to the context, without fabricating a prototype.
            While this serves as a preview and a coarse validation, it does not provide a holistic model of all interactions between the object and the context, which would require a maximum degree of fabrication-awareness and scene understanding.
            While this is not provided in our proof-of-concept implementation of \system yet, it is likely to be feasible with future extended reality platforms, which, by design, rely on scene understanding to function.  \section{Limitations}\label{sec:limitations}
    Our research exhibits a set of limitations, which we discuss in the following paragraphs.
       
    \paragraph{Pipeline}
        The pipeline in \system (cf. \autoref{fig:process}) assumes that the fitting customizable object has already been found. 
        However, for most users, the process starts with searching objects, which is a non-trivial task and has been considered in works like CustomizAR~\cite{liangCustomizARFacilitatingInteractive2022} or ShapeFindAR~\cite{stemasovShapeFindARExploringInSitu2022}.
        \system's pipeline also ends before the actual acquisition step, which we deemed to be optimizable without user interference (e.g., considering 3D print services over fabrication at home).
        Nevertheless, we completed every step of the process from design configuration to fabrication, outlined in \autoref{fig:process}, ourselves several times.
        This includes multiple 3D printed lampshades (e.g., \autoref{fig:teaser}.4, top and \autoref{fig:lighting}) and a CNC-routed wooden shelf (\autoref{fig:teaser}.4, bottom). 
        We consider this to be an instance of ``Evaluation by Demonstration''~\cite{ledoEvaluationStrategiesHCI2018}, albeit a compact one.
        The lampshades were printed using an Ultimaker 2+ 3D printer\fn{https://ultimaker.com/3d-printers/s-series/ultimaker-2-connect/}{24.01.2024}, and the shelf was milled using a Shaper Origin\fn{https://www.shapertools.com/en-us/origin}{25.01.2024}. 
        Both required the use of CAM\footnote{For the lampshades: Ultimaker Cura (\url{https://ultimaker.com/software/ultimaker-cura/}, Accessed: 24.01.2024), and Slicer for Fusion 360 (\url{https://www.autodesk.com/support/technical/article/caas/tsarticles/ts/3yg7zznS94MHNDG7KMV8Qg.html}, Accessed: 25.01.2024) for the shelf.} (computer-aided manufacturing) software, which could be integrated into the tool, as demonstrated by works like Mix\&Match~\cite{stemasovMixMatchOmitting2020}, which included a slicer.
        Including this step in future evaluations is promising for understanding the breakdowns in transferring a user's design to a manufactured artifact.

    \paragraph{Object Set}
        \system and, even more so, the study, focused on a limited set of designs to customize.
        While they cover a broad range of scenarios and parameter types, an even richer set of parametric designs may uncover different interaction patterns or the need for other validation functionalities.
        This particularly applies to artifact classes beyond established machines for personal fabrication (i.e., 3D-Printers, CNC mills, laser cutters).
        For example, parameters for clothing are likely more complex to measure and incorporate in a design~\cite{saakesMirrorMirrorOnbody2015,wibowoDressUp3DInterface2012}, as they are potentially more organic and irregular, yet uniquely personal.

    \paragraph{Authoring}
        As outlined in \autoref{par:definitions}, our focus lies on the \emph{customization} of parametric design through end-users.
        However, there is also demand for such designs to be \emph{created} and \emph{authored} by expert designers~\cite{carpoSecondDigitalTurn2017,flathCopyTransformCombine2017} to enable customization in the first place (i.e., through meta-design~\cite{fischerMetadesignDesignDesigners2000}).
        More complex environments (e.g., Archimatix or Grasshopper) are appropriate here, and could benefit from in-situ approaches nonetheless.
        With approaches like \system, authoring customizable designs ought to change in parallel, as designers have to encode more aspects in their parametric design than before (i.e., more than static geometry).
        For \system, this includes all relations to ergonomic properties (e.g., what influences the seating area or where an arm may rest), or individual handles, their positions, and their direction indicators (cf. \autoref{fig:directManip}.2).
        This can also include configuring multiple lights instead of a point light only, or including other simulation domains, like fluids or aerodynamics, and encoding their relationship with the context accordingly.

    \paragraph{Estimations $\neq$ Simulations}
        We consciously label the evaluation functions of \system as \textit{``estimations''}, not \textit{``simulations''}.
        They require little configuration by the user and deliberately do not provide a high fidelity.
        Nevertheless, we consider the results of the estimations (balance, lighting) to be a way of early, worst-case evaluation~\cite{stemasovBrickStARtEnablingInsitu2023}, which, unlike others, takes into account the context (e.g., a surface an object will rest on).
        With increasingly detailed scene understanding coupled with a high degree of fabrication awareness, these estimations will likely increase in value without added complexity for the end users.
        
    \paragraph{Study Design and Implementation Platform}
Despite both consisting of subtasks that cover both ends of the spectrum between form and function, participants did not perceive the tasks as similarly complex. 
        While fundamental differences and benefits between \system and its desktop counterpart were confirmed, further studies to understand \system as a design- and creativity-support-tool are needed, ideally covering the entire pipeline from design choice to fabrication.
        Furthermore, despite our goal to achieve full feature-parity between \system and the desktop variant, the curve drawing feature was absent from the desktop, which may explain the high usefulness rankings (\autoref{sec:prefs}).
Most importantly, the ergonomic issues (e.g., field of view, interaction precision) were not ideal and likely influenced participants' assessments.
        This also applies to the fidelity of the environment scan that informed the light and balance estimations.

    \paragraph{Interface and Underlying Constraints}
        The fundamental elements of the UI can also be expanded, as both the interactions and the gizmos used (e.g., handles for direct manipulation or sliders on the UI) were deliberately simple.
        While this worked well for the user study, there is room for improvement and more fine-grained explorations (cf. \cite{dreyVRSketchInExploringDesign2020,reipschlagerDesignARImmersive3DModeling2019,kimTangible3DHand2005}).
        This also applies to the detailed implementation of constraints, where, for example, an invalid curve yields a valid result that may not resemble the user's input closely.\\

    Despite these limitations, we are confident that the in-situ interaction with parametric design, as demonstrated by \system, is a meaningful and highly promising augmentation of how we currently use parametric designs in the context of personal fabrication.
 \section{Future Work}\label{sec:futurework}
    \system opens up several avenues for future work, related to added functionalities for \system, further user studies, but also to the concept of \concept itself.
    
\paragraph{Richer Model Set} Similarly to prior art (e.g.,~\cite{shugrinaFabFormsCustomizable2015}), \system operates on a limited set of pre-designed models (cf. \autoref{fig:models}) that can certainly be expanded.
    This would require feasible search functionalities \cite{buschelHereNowRealityBased2018,stemasovShapeFindARExploringInSitu2022}. This richer set of models likely opens up new opportunities for interaction, like using a 3D scan -- afforded by the use of an XR headset~\cite{stemasovMixMatchOmitting2020} -- as a higher-level input to a parametric design.
    An early prototype of \system also used OpenPose to automatically acquire ergonomics-related body measures -- similarly to Kim et al.~\cite{kimErgonomicPosturalAssessment2021} -- but was found to be unreliable and hard to reconcile with a headset as a platform.
    The limitations in \autoref{sec:limitations} can also inform future directions, such as improving feedback and transparency regarding constraints and the logic they may follow to the user, as the current implementation of \system focused on exploring individual interaction techniques.

    \paragraph{Generalizing beyond Fabrication} 
    In line with expanding the model set would be the consideration of domains beyond (personal) fabrication.
    Physical context and interactions with this environment are crucial to consider prior to \emph{fabrication}, but can also inform purely digital domains where parametric designs can be relevant (e.g., digital design~\cite{michelDAGAmendmentInverse2021}).
    Here, understanding requirements of non-novices and scalability to more complex designs pose highly intriguing challenges.

    \paragraph{Automation and Evaluations} 
    We also aim to further explore automation in this context, to further benefit from these more abstract expressions of parameters: one may consider users starting with vague gestural \cite{leePosingActingInput2016,holzDataMimingInferring2011}, spatial \cite{stemasovShapeFindARExploringInSitu2022}, textual, or multimodal \cite{fraserReMapLoweringBarrier2020} descriptions of requirements to receive fitting results from a customizable design.
    The estimation components can also be improved (i.e., made more precise) and expanded, through approaches like fluid simulations (e.g., for the design of hydroponics systems~\cite{pearceApplicationsOpenSource2015, takeuchi3DPrintableHydroponics2019}), or load-bearing simulations~\cite{abdullahFastForceRealTimeReinforcement2021}, to further focus on functional mechanical parts.
    Balancing this degree of early evaluation with the tool's complexity is non-trivial and additionally demands a higher degree of \emph{fabrication-awareness} (e.g., how dense the used material is).
    Lastly, we plan user studies to understand usage patterns further and optimize for interface ergonomics.  \section{Conclusion}\label{sec:conclusion}
Finding and starting with a finished design is a highly beneficial approach to 3D-modeling, but is limited in expressivity.
    Parametric designs lower the effort needed to engage in 3D-design by replacing detailed modeling with configuration.
    This notion is usually ex-situ and currently benefits domain experts (e.g., architects), or enthusiasts configuring select 3D models on repositories like \tv.
    
To augment in-situ search-and-fabricate approaches and further simplify interactions with parametric designs, we presented the concept of \concept and introduced \system, a proof-of-concept implementation embodying this notion.
    \system leverages extended reality to 1) support users in choosing parameter values (e.g., through gesture-based inputs or recommendations), and 2) provide early, in-context estimation functionalities (e.g., lighting).
    We evaluated \system in a user study and found that it allows users to incorporate context into their design process while also benefiting from in-situ estimations.

We argue that extended reality, as an emerging technology, is a crucial support component to design for (personal) fabrication, as it enables not only in-situ interaction, but also provides context understanding that can be woven into design tools themselves, as demonstrated with \system.
    This opens up a rich space for design support, relevant for parametric designs and other approaches to 3D-design by bridging unique user requirements, their unique contexts, and their representations in design tools.
    Leveraging this combination may make design for personal fabrication even more accessible to end-users, without demanding physical design iterations or complex information transfers. 
     
\begin{acks}
    We thank the participants of our user study for their time and the reviewers of the manuscript for their thoughtful and helpful feedback.\\

    This research was partially funded by the Deutsche Forschungsgemeinschaft (DFG, German Research Foundation) through the project ``\textit{Democratic and Sustainable Personal Design and Fabrication through In-situ Co-Design and Previsualization}'' (Project number: 525038300).\\

    This work has been co-funded by the LOEWE initiative (Hesse, Germany) within the emergenCITY center.
\end{acks}

\bibliographystyle{ACM-Reference-Format}
\bibliography{zotero.bib}

\appendix
\pagebreak
\section{Additional Screenshots}\label{sec:appendix-a}
            \begin{figure}[h!]
                \centering
                \includegraphics[width=\minof{\columnwidth}{0.65\textwidth}]{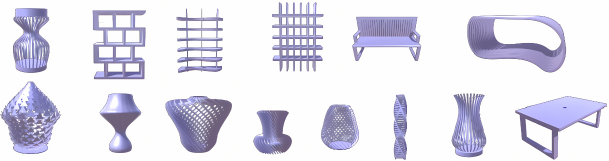}
                \caption{Parametrized models implemented for and included in our proof-of-concept implementation of \system. The selection contains a variety of vases/lampshades, shelf models, tables, and a bench. Not pictured: the book holder used as a training task in the study (cf. \autoref{fig:desktopVsAr}b).}
                \label{fig:models}
            \end{figure}

        \begin{figure}[h!]
                \centering
                \includegraphics[width=\minof{\columnwidth}{0.65\textwidth}]{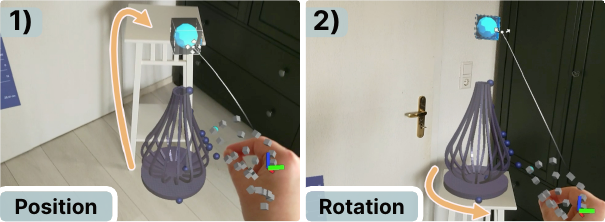}
                \caption{A gizmo (cyan polyhedron), as implemented as a default in the MRTK for object manipulation, is positioned above the model and can be used for positioning (1) and rotation (2) of the model.}
                \label{fig:positionAndRotation}
            \end{figure}
        
            \begin{figure}[h!]
                \centering
                \includegraphics[width=\minof{\columnwidth}{0.65\textwidth}]{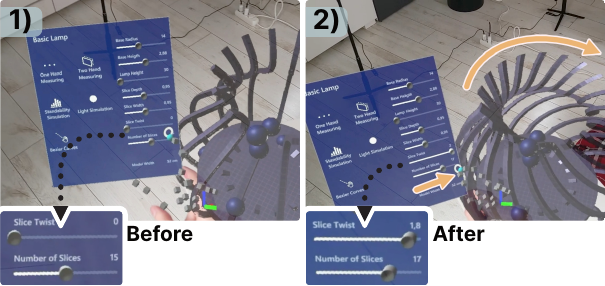}
                \caption{\system depicts a list of parameters on a floating UI for the user. A pinch gesture lets users manipulate values through sliders in the UI (1). The changes are immediately reflected in an updated model preview (2). Here, a user changed the \textit{slice twist} parameter from 0 to 1.8.}
                \label{fig:uiManip}
            \end{figure}
            
\clearpage
\section{Participant Interactions}\label{sec:appendix-b}\begin{figure*}[b!]
    \centering
\includegraphics[width=0.94\textwidth]{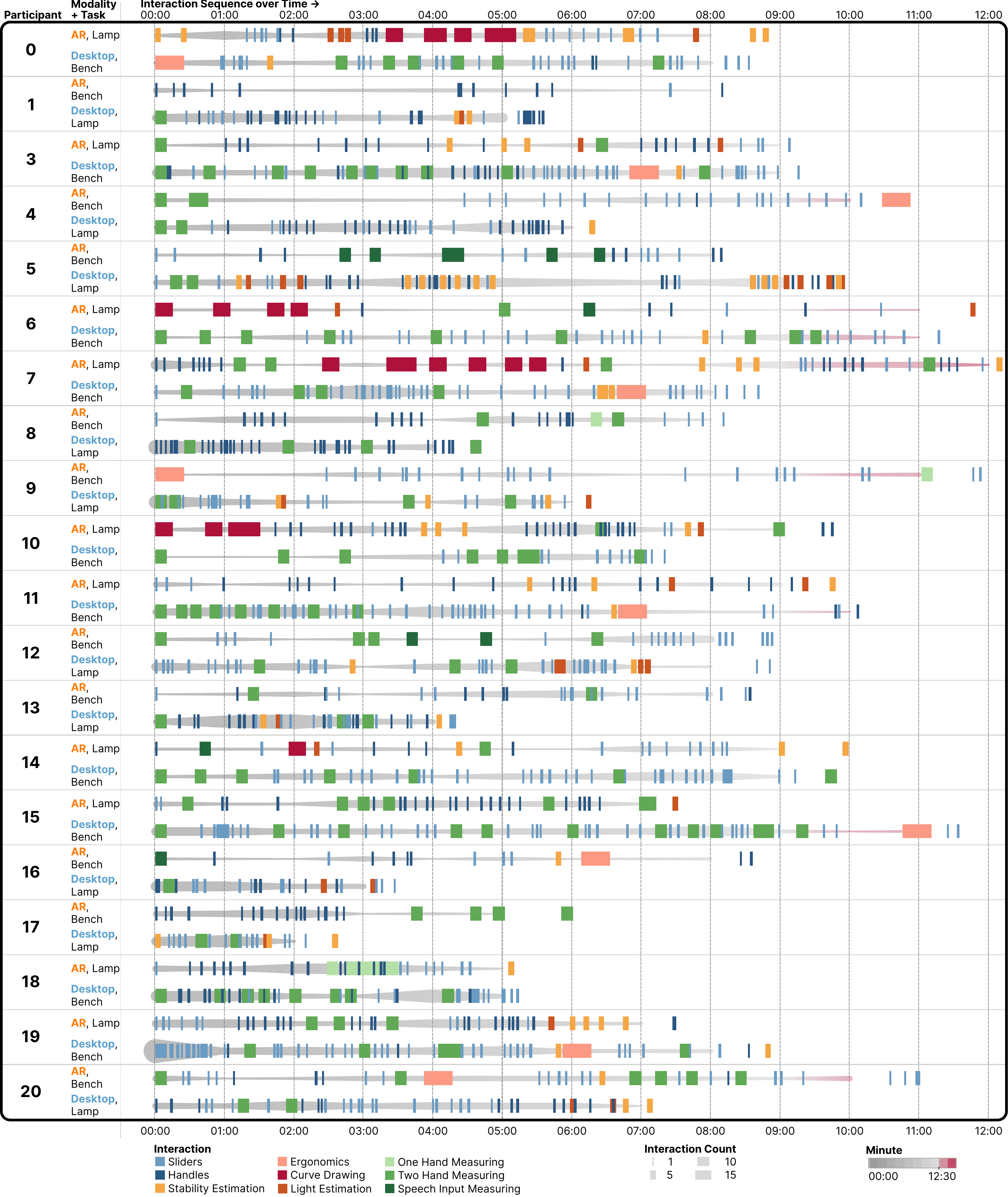}
    \caption{Interaction sequences for each study participant, with tool-task combinations annotated in the second column. Note: P2 had to drop out, and the replacement is labeled with ID 0 in this plot and ID 21 across the rest of the paper.}
    \label{fig:interactionSequencePerUser}
\end{figure*} 
\end{document}